\documentclass[pra,twocolumn,superscriptaddress,preprintnumbers,amsmath,amssymb]{revtex4}
\usepackage{ }

\usepackage{color}
\usepackage[colorlinks=true]{hyperref}
\hypersetup{colorlinks=true,linkcolor=blue,citecolor=blue,urlcolor=blue}
\usepackage{bm}
\usepackage{bbm}
\usepackage{amssymb}
\usepackage{amsfonts}
\usepackage{epsfig,graphicx}
\usepackage{amstext}
\usepackage{amsmath}
\usepackage{graphicx}
\usepackage{times}
\usepackage{txfonts}
\usepackage{dcolumn}
\usepackage{dsfont}
\usepackage{braket}

\newcommand{\tr}{\mathrm{tr}}

\begin{document}

\title{Efficient wireless charging of a quantum battery}

\author{Ming-Liang Hu}
\email{mingliang0301@163.com}
\address{School of Science, Xi'an University of Posts and Telecommunications, Xi'an 710121, China}

\author{Ting Gao}
\address{School of Science, Xi'an University of Posts and Telecommunications, Xi'an 710121, China}

\author{Heng Fan}
\email{hfan@iphy.ac.cn}
\affiliation{Institute of Physics, Chinese Academy of Sciences, Beijing 100190, China}
\affiliation{School of Physical Sciences, University of Chinese Academy of Sciences, Beijing 100190, China}
\affiliation{Beijing Academy of Quantum Information Sciences, Beijing 100193, China}

\begin{abstract}
We explore the wireless charging of a quantum battery (QB) via $n$ charging units, whose coupling is mediated by a common bosonic reservoir. We consider the general scenarios in which the charger energy is not maximal and the QB has residual ergotropy initially. It is found that the charging performance improves with the increase of the coupling strength. In the strong coupling regime, the charging time is insensitive to the charger energy, the number of charging units, and the residual ergotropy in the QB, while the ergotropy charged on the QB strongly depends on the charger energy and ergotropy, and the residual ergotropy in the QB does not help to enhance its performance. Moreover, the multiple charging units help to enhance the charging performance in the weak and moderate coupling regimes, while they are less efficient in the strong coupling regime.
\end{abstract}


\maketitle

\section{Introduction} \label{sec:1}
Quantum battery (QB) is a microscopic quantum mechanical device used to temporarily store and release energy in a controllable manner. The idealized procedure of work extraction from a QB is implemented via a cyclic unitary transformation of it governed by the system dynamics plus some controlling fields in a certain time interval. The maximal amount of extractable work in this process is termed as ergotropy \cite{EPL}, which is defined as the surplus energy of the initial state $\rho$ of the QB with respect to the full set of passive states having the same eigenvalues as $\rho$. Here, a state $\sigma$ is called passive if no work can be extracted from it by using any such unitary transformation \cite{passive1,passive2,passive3}, while it is called active otherwise. Moreover, the product of two independent copies of a passive state $\sigma$ may not always be passive ($\sigma$ is called completely passive if $\sigma^{\otimes n}$ is passive with respect to the sum Hamiltonian $H=\sum_{i=1}^n H_i$, $\forall n$), so the ergotropy for $n$ collective QB cells may be different from the sum of ergotropies for $n$ copies of a single-cell QB \cite{EPL,entangleu}.

To date, the QB has been extensively studied from two main aspects. The first one, centering around the interplay between the performance of a QB and the various quantum resources, revealed that some of its figures of merit  (ergotropy, charging power, battery capacity, etc.) are intimately related to quantum characteristics of the battery state such as entropy, coherence, and entanglement \cite{workext1,workext2,qcqb0,qcqb1,qcqb2,qcqb3,qcqb4,qcqb5, shihl1,qcqb6,correch,shihl2,luomx,shihl3,AVS}. The second one, on the other hand, focused on the charging process of a QB, e.g., the possible realizations of a QB in the spin-chain systems \cite{shihl1,spin_qb1,spin_qb2, spin_qb3,spin_qb4,spin_qb5}, the Tavis-Cummings model \cite{qcqb1,TCqb1}, the harmonic oscillators \cite{noiseqb0,oscillator_qb1,oscillator_qb2}, the Dicke model \cite{Dickeqb1,Dickeqb2,Dickeqb3,Dickeqb4}, and the three-level systems \cite{oscillator_qb2,threelevelqb,Dou1,Dou2}. For these battery models, numerous efforts have also been invested into devising feasible schemes to enhance their charging power \cite{chpf0,chpf1, thermalization,chpf2,chpf3, chpf4,chpf5,chpf6,chpf7,Shangc}, exploring their quantum advantages in the charging efficiency (time of charging, input power, total stored energy, etc.) \cite{advan1,advan2,SYKbattery,advan3, advan4,advan5,advan6,advan7}, and identifying bounds on their extractable work and charging power \cite{qcqb2,bound1,bound2}. Moreover, as any quantum system is inevitably disturbed by its surrounding environment, and the charger and QB cannot be excluded from this process too, the dissipative charging of a QB has also been widely explored \cite{qcqb4,noiseqb0,noiseqb1,wireless1,wireless2,noiseqb2,PRAppl,noiseqb3,noiseqb4,noiseqb5,noiseqb6,noiseqb7, collective,noiseqb8, squeezqb, twomode,anjh,weak}; in particular, some structured reservoirs can be exploited as mediators for transferring energy from the charger to a QB, that is, to realize the wireless charging process \cite{wireless1,wireless2,twomode}.

Concerning the charging of a QB, our goal is to realize the fast charging and high charging capacity, and it is appealing to provide a general study on the details of the charging process under various situations. However, while most of the previous works considered the case where the charger energy is maximal and the QB is fully discharged initially, the charger might not always be in its fully excited state and one rarely waits until a battery runs out before charging in realistic situations. In fact, by considering different initial states of a QB driven by a classical force, it was found that an optimal charging is achieved if it starts from the ground state \cite{initial}. Inspired by this, we consider in this paper the general scenarios in which the charger energy is not maximal and the QB is not fully discharged initially, putting emphasis on the role of the number $n$ of charging units and the amount of energy in the charger, the residual ergotropy in the QB, and the coupling strength of the system (i.e., charger $+$ QB) to a common structured reservoir in controlling the charging time and the ergotropy. We will show that the charging performance mediated by the common reservoir improves with the increasing coupling strength, and counter intuitively, the multiple charging units and the residual ergotropy in the QB may not always be beneficial for shortening the charging time and enhancing the charged ergotropy.

The rest of the paper is organized as follows. In Sec. \ref{sec:2} we recall the concept of ergotropy, and in Sec. \ref{sec:3} we present the charger-battery model we considered. Section \ref{sec:4} is devoted to analyzing the charging of a QB via $n$ ($n\geqslant 1$) charging units. Finally, we summarize our findings in Sec. \ref{sec:5}.

\section{Preliminaries} \label{sec:2}
In this section, we recall the notion of ergotropy and the related quantities. The ergotropy quantifies the maximal amount of work that can be reversibly extracted from a system (i.e., a QB). For a given QB with the free Hamiltonian $H$, the average work extractable from it by using a cyclic unitary transformation $U$ is $\mathcal{W}(\rho,U)= \tr(\rho H)- \tr(U\rho U^\dag H)$, and the ergotropy $\mathcal{E}$ is defined as the maximum of $\mathcal{W}$, i.e.,
\begin{equation}\label{eq2-1}
 \mathcal{E}= \max_{U\in \mathcal{U}_c} \mathcal{W}(\rho,U),
\end{equation}
where $\rho$ is the battery state and the maximization is taken over the set of cyclic unitary operations $\mathcal{U}_c$ in the Hilbert space $\mathcal{H}_d$ ($d=\dim\rho$), and $\mathcal{U}_c$ can be generated in a given time interval $[0,\tau]$ by applying suitable control fields to the system.

The optimal state $\tilde{\rho}= \tilde{U}\rho\tilde{U}^\dag$, realizing the maximum in Eq. \eqref{eq2-1}, is called the passive state associated with $\rho$. By rewriting $\rho$ in its eigenbasis as $\rho= \sum_j r_j |r_j\rangle \langle r_j|$ and $H$ in its eigenbasis as $H= \sum_k \varepsilon_k |\varepsilon_k\rangle \langle \varepsilon_k|$, where their respective eigenvalues are reordered as $r_j \geqslant r_{j+1}$ ($\forall j$) and $\varepsilon_k \leqslant \varepsilon_{k+1}$ ($\forall k$), the optimal unitary can be obtained as $\tilde{U}=\sum_j |\varepsilon_j\rangle \langle r_j|$, and the corresponding passive state is $\tilde{\rho}= \sum_j r_j |\varepsilon_j\rangle \langle \varepsilon_j|$. The ergotropy $\mathcal{E}$ can then be obtained explicitly as \cite{EPL}
\begin{equation}\label{eq2-2}
\mathcal{E}=  \sum_k \varepsilon_k (\rho_{kk}-r_k),
\end{equation}
where $\rho_{kk}= \sum_j r_j |\langle r_j|\varepsilon_k\rangle|^2$ is the $k$th diagonal element of $\rho$ in the energy eigenbasis $\{|\varepsilon_k\rangle\}$ of $H$.

The ergotropy $\mathcal{E}$ can be divided into two components, i.e., the incoherent and coherent components \cite{qcqb3}. The incoherent component $\mathcal{E}_i$ corresponds to the maximum work that can be extracted from the QB under coherence preserving operations $\mathcal{U}_c^{(i)}$, i.e.,
\begin{equation}\label{eq2-3}
 \mathcal{E}_i= \max_{V\in \mathcal{U}_c^{(i)}} \mathcal{W}(\rho,V),
\end{equation}
where $V=\sum_k e^{-i\phi_k}|\varepsilon_k\rangle \langle \varepsilon_{\pi_k}| \equiv V_\pi$, with $\phi_k$ being an irrelevant phase factor and $\{\pi_k\}_{k=1,\dots, d}$ being a permutation of the elements of $\{1,\dots,d\}$. After optimizing over all possible $\{\pi_k\}_{k=1,\dots, d}$, one can obtain that the optimal $V_{\tilde{\pi}}$ ($\tilde{\pi}$ is the optimal permutation) yields $\sigma_\rho=V_{\tilde{\pi}}\rho V_{\tilde{\pi}}^\dag= \sum_{kl} \rho_{\tilde{\pi}_k \tilde{\pi}_l}|\varepsilon_k\rangle\langle\varepsilon_l|$ \cite{qcqb3}, thus
\begin{equation}\label{eq2-4}
\mathcal{E}_i=  \sum_k \varepsilon_k (\rho_{kk}-\rho_{\tilde{\pi}_k \tilde{\pi}_k}),
\end{equation}
where $\{\rho_{\tilde{\pi}_k \tilde{\pi}_k}\}_{k=1,\dots, d}$ is the rearrangement of the populations $\{\rho_{kk}\}_{k=1,\dots, d}$ in decreasing order. Equation \eqref{eq2-4} indicates that $\mathcal{E}_i$ also equals the ergotropy of the dephased state $\delta_\rho= \Delta[\rho]=\mathrm{diag}\{\rho_{11}, \ldots, \rho_{dd}\}$, i.e., $\mathcal{E}_i(\rho)=\mathcal{E}(\delta_\rho)$ \cite{shihl2}, and the passive state associated with $\delta_\rho$ is $\tilde{\delta}_\rho=\mathrm{diag} \{\rho_{\tilde{\pi}_1 \tilde{\pi}_1}, \dots, \rho_{\tilde{\pi}_d \tilde{\pi}_d}\}$.

Having defined the incoherent component $\mathcal{E}_i$, the coherent component of $\mathcal{E}$ is naturally defined as $\mathcal{E}_c= \mathcal{E}-\mathcal{E}_i$, which, by combining Eqs. \eqref{eq2-2} and \eqref{eq2-4}, can be obtained as
\begin{equation}\label{eq2-5}
\mathcal{E}_c= \sum_k \varepsilon_k  (\rho_{\tilde{\pi}_k \tilde{\pi}_k}-r_k),
\end{equation}
and it quantifies the amount of work which cannot be extracted by using only incoherent operations.

\section{The charger-battery model} \label{sec:3}
The charger-battery model that we are going to discuss contains $N$ qubits, all of which are placed inside a common zero-temperature bosonic reservoir in the vacuum initially. The total Hamiltonian (in units of $\hbar$), in the rotating wave approximation, can be written as
\begin{equation}\label{eq3-1}
 H=H_S+H_R+H_{\mathrm{int}},
\end{equation}
where $H_S$ and $H_R$, describe, respectively, the free Hamiltonians of the system and the reservoir, while $H_{\mathrm{int}}$ describes the interaction of the system with the reservoir. Their explicit forms are as follows:
\begin{equation}\label{eq3-2}
\begin{aligned}
 & H_S= \omega_0 \sum_i \sigma_i^{+} \sigma_i^{-},~
   H_R= \sum_k \omega_k b_k^\dagger b_k, \\
 & H_{\mathrm{int}}= f(t)\sum_{i,k} g_k\sigma_i^{+}b_k + \mathrm{H.c.},
\end{aligned}
\end{equation}
where $\omega_0$ is the transition frequency between the ground state $|0\rangle$ and excited state $|1\rangle$ of each qubit and $\omega_0 \sigma_i^{+} \sigma_i^{-}$ is actually the free Hamiltonian of the $i$th qubit, with $\sigma_i^{+}$ ($\sigma_i^{-}$) being the Pauli raising (lowering) operator. Moreover, $b_k$ ($b_k^\dagger$) is the annihilation (creation) operator of the field mode $k$ with frequency $\omega_k$, and $g_k$ is the coupling strength between each qubit and the field mode $k$ of the reservoir. The function $f(t)$, which equals 1 for $t\in (0,\bar{t}]$ and 0 otherwise, is introduced for controlling the switchable coupling of the $N$ qubits to the reservoir, where $\bar{t}$ is the interaction time needed to charge the QB up to its (not necessarily the first) dynamical maximum. When $t>\bar{t}$, the interaction between the qubits and the reservoir is switched off. Hereafter we call $\bar{t}$ the charging time.

By treating the first $n$ qubits as the charger and the other $N-n$ qubits the QB, the charging process can be implemented as follows. First, the ``charger $+$ battery" system is prepared in the state $\rho_S$, and its interaction with the reservoir is switched off at $t\leqslant 0$, that is, the initial state of the $N$-qubit system and the reservoir is $\rho=\rho_S \otimes |\bar{0}\rangle\langle \bar{0}|$, where $|\bar{0}\rangle$ is the vacuum state of the reservoir. When $t > 0$, the interaction between the $N$-qubit system and the reservoir is switched on, thereby there will be reservoir-mediated indirect interactions among the qubits, due to which the energy in the charger can be transferred to the QB in a wireless manner, and the common reservoir plays the role of a mediator between the two elements.

To elucidate the above wireless charging process, we need to solve the evolution equation of the $N$ qubits. In this paper, we focus on the case in which the reservoir is the electromagnetic field inside a lossy cavity, which displays a Lorentzian broadening due to the nonperfect reflectivity of the cavity mirrors, and the spectral density is given by \cite{common1}
\begin{equation}\label{eq3-3}
 J(\omega)= \frac{\Omega^2}{\pi} \frac{\lambda}{(\omega-\omega_0)^2+\lambda^2},
\end{equation}
where $\Omega$ is the effective coupling strength proportional to the vacuum Rabi frequency, $\lambda$ denotes the frequency width of the spectrum, and the ideal cavity limit (no losses) is obtained at $\lambda \rightarrow 0$. For this model, if $N=2$ (i.e., the single charger case) and the ``charger $+$ battery" system is initialized in $v_{01}|10\rangle+ v_{02}|01\rangle$ ($|v_{01}|^2+|v_{02}|^2=1$), the dynamics of the system can be obtained analytically \cite{common1,common2}, and the associated QB has been analyzed in Refs. \cite{wireless1,wireless2}. For the initial extended Werner-like states, analytical solutions of the two-qubit system can also be obtained in the Laplace transform space \cite{pmode0}.

For a general initial state, it is hard to obtain the exact dynamics of the $N$-qubit system, even for $N=2$. So, we resort to the pseudomode approach, which describes the coherent interaction between the considered system and the pseudomodes \cite{pmode1,pmode2,pmode3}. Here, the pseudomodes are auxiliary variables defined from the spectrum of the reservoir. For $J(\omega)$ of Eq. \eqref{eq3-3}, this approach results in the following pseudomode master equation in the interaction picture \cite{pmode1,pmode2,pmode3,pmode4,pmode5}:
\begin{equation}\label{eq3-4}
\frac{\partial \varrho}{\partial t}= -i[V, \varrho] +\lambda (2 a\varrho a^\dagger- a^\dagger a \varrho - \varrho a^\dagger a),
\end{equation}
where $a$ ($a^\dagger$) is the annihilation (creation) operator of the pseudomode, $\varrho$ is the density operator of the extended system comprising the $N$-qubit system and the pseudomode, while the effective coupling between the $N$-qubit system and the pseudomode is described by the interaction Hamiltonian
\begin{equation}\label{eq3-5}
 V=\Omega\sum_i \sigma_i^{+} a+ \mathrm{H.c.};
\end{equation}
by combining this with Eq. \eqref{eq3-4}, one can solve numerically the dynamics of $\varrho$ without performing any further approximation. After having $\varrho$ at hand, the density operator $\rho$ for the $N$-qubit system can then be obtained via partial tracing of the pseudomode degrees of freedom, and likewise for the charger state $\rho_{\mathrm{ch}}$ and the battery state $\rho_{\mathrm{ba}}$.

In alignment with Refs. \cite{common1,common2,wireless1}, hereafter we use the dimensionless parameter $R=\sqrt{2}\Omega/\lambda$ to distinguish the strong coupling regime (good cavity, $R\gg 1$) from the weak one (bad cavity, $R \ll 1$), and due to the limited computing resource, we will focus on the cases of $N=2$, 3, and 4, respectively.

\section{Charging the QB with different chargers} \label{sec:4}
\begin{figure}
\centering
\resizebox{0.46 \textwidth}{!}{%
\includegraphics{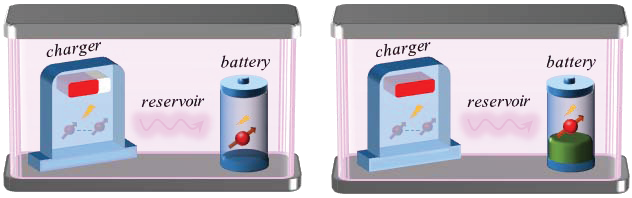}}
\caption{Schematic representation of the wireless charging of the QB. (Left) Scenario \uppercase\expandafter{\romannumeral+1}, where the charger is not in its excited state and the QB is empty initially. (Right) Scenario \uppercase\expandafter{\romannumeral+2}, where the charger is in its excited state and the QB has residual ergotropy initially.} \label{fig:1}
\end{figure}

In this work, we consider the wireless charging of a QB for two different scenarios. For convenience of later presentation, we term them as scenarios \uppercase\expandafter{\romannumeral+1} and \uppercase\expandafter{\romannumeral+2}, respectively. As sketched in Fig. \ref{fig:1}, scenario \uppercase\expandafter{\romannumeral+1} refers to the case in which the charger is in the state $|\psi\rangle_{c_1 \ldots c_n}=\otimes_{i=1}^{n} |\psi\rangle_{c_i}$ and the QB is in the ground state $|0\rangle$ initially, while scenario \uppercase\expandafter{\romannumeral+2} refers to the case in which the charger is in the fully excited state $|1\rangle^{\otimes n}$ and the QB is in the active state $|\varphi\rangle_{e_1}$ initially. The forms of $|\psi\rangle_{c_i}$ ($i=1,\dots,n$) and $|\varphi\rangle_{e_1}$ are as follows:
\begin{equation}\label{eq4-1}
\begin{aligned}
 & |\psi\rangle_{c_i}= \sqrt{c_i}|1\rangle + \sqrt{1-c_i} |0\rangle ~ (i=1,\dots,n), \\
 & |\varphi\rangle_{e_1}= \sqrt{e_1}|1\rangle + \sqrt{1-e_1} |0\rangle,~
\end{aligned}
\end{equation}
where the parameters $c_i\in[0,1]$ and $e_1\in[0,1]$.

For scenario \uppercase\expandafter{\romannumeral+1}, the $n$ charging units are not in their excited states, that is, the charger energy is not maximal initially. So, for the case of $n=1$, even if the total energy in the charger is transferred to a QB, it cannot be fully charged. But, this is a common situation encountered in practice, and it motivates us to consider the charging of an empty QB via $n$ charging units having nonmaximal energy, aimed at exploring whether it can complement the lack of energy in a single charging unit. Here, by saying a QB is empty, we mean that no work can be extracted from it via any cyclic unitary transformation, that is, the battery state is passive. But, this does not necessarily mean it is in the ground state. For scenario \uppercase\expandafter{\romannumeral+2}, there is residual ergotropy in the QB (i.e., $\mathcal{E}\neq 0$ initially), which is also a situation we may encounter, as we rarely wait until a battery runs out before charging. In particular, a traditional battery having residual energy may be charged faster than a fully discharged one. But is this indeed the case for a QB? This is our motivation for considering scenario \uppercase\expandafter{\romannumeral+2}.

Apart from the initial states in Eq. \eqref{eq4-1}, we will also mention some other states (see below), e.g., the correlated charger states and mixed battery states.

\subsection{The case of a single charging unit} \label{sec:4a}
\begin{figure}
\centering
\resizebox{0.45 \textwidth}{!}{%
\includegraphics{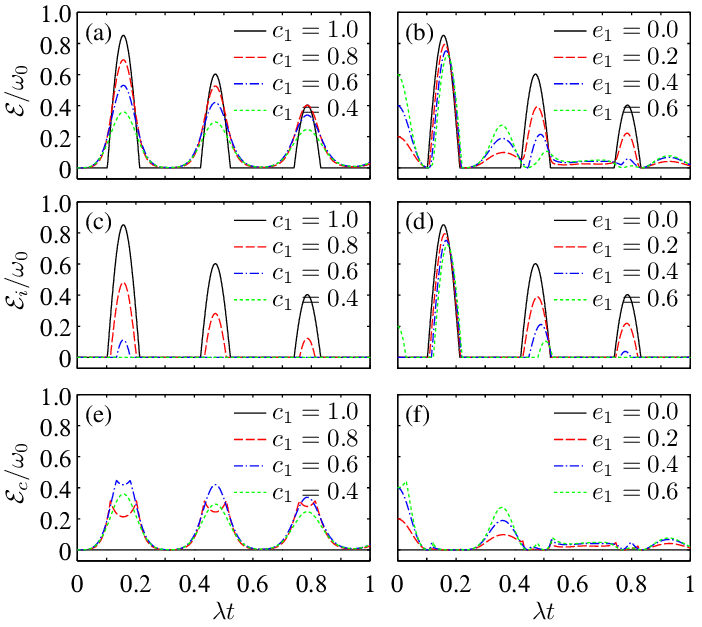}}
\caption{Dynamics of $\mathcal{E}$, $\mathcal{E}_i$, and $\mathcal{E}_c$ of the QB in the good cavity limit $R=20$ for the case of a single charging unit. The left (right) three panels are plotted with the charger and QB being in the initial states $|\psi\rangle_{c_1}$ and $|0\rangle$ ($|1\rangle$ and $|\varphi\rangle_{e_1}$), respectively.} \label{fig:2}
\end{figure}

For the case of a single charging unit (i.e., $n=1$), if $c_1$ in Eq. \eqref{eq4-1} is equal to 1, the charging of a QB has been investigated in Refs. \cite{wireless1,wireless2}. For scenario \uppercase\expandafter{\romannumeral+1} with different initial charger states $|\psi\rangle_{c_1}$, we show in the left column of Fig. \ref{fig:2} the dynamics of the ergotropy $\mathcal{E}$ as well as its incoherent component $\mathcal{E}_i$ and coherent component $\mathcal{E}_c$ (in units of $\omega_0$) in the good cavity limit $R=20$. For this charger, its initial energy is $c_1\omega_0$, and the first dynamical maximum of $\mathcal{E}$ corresponds to the maximal ergotropy charged on the QB, which, as shown by the different lines in Fig. \ref{fig:2}(a), decreases with a decrease in $c_1$. This is understandable as the charger energy also decreases with a decrease in $c_1$. Moreover, our calculation shows that the energy is transferred to the QB from the very beginning, but as is shown in Fig. \ref{fig:2}(a), the ergotropy $\mathcal{E}$ remains 0 for a finite time interval ($\lambda t\lesssim 0.1023$ when $R=20$). This shows that the transferred energy cannot be converted into extractable work immediately in certain situations. As for the incoherent and coherent contributions to the ergotropy, by comparing the left three panels of Fig. \ref{fig:2}, one can see that $\mathcal{E}_c \equiv 0$ for the charger with $c_1=1$. Moreover, if $0.7073 \lesssim c_1<1$, the incoherent and coherent components of the ergotropy dominate its dynamics alternately, and if $c_1 \lesssim 0.7073$, the coherent contribution will always be dominant.

When considering scenario \uppercase\expandafter{\romannumeral+2}, we show in the right column of Fig. \ref{fig:2} the dynamics of $\mathcal{E}$, $\mathcal{E}_i$, and $\mathcal{E}_c$ (in units of $\omega_0$) in the good cavity limit $R=20$. For this scenario, the initial state $|\varphi\rangle_{e_1}$ of the QB is active if $e_1 \neq 0$ and the residual ergotropy in it is $e_1\omega_0$. From Fig. \ref{fig:2}(b) one can see that $\mathcal{E}$ first decays to 0 (i.e., fully discharged) and then turns to be increased to its first dynamical maximum, and this can be recognized as the maximal ergotropy charged on the QB as the subsequent dynamical maxima become smaller and smaller. This phenomenon is in contrast to that of a traditional battery, as it indicates that when there is residual ergotropy in the QB, it will inevitably undergo a fully discharging process, after which it can then be continuously charged. Moreover, for a fixed $R$, the first dynamical maximum of $\mathcal{E}$ may be smaller than its initial value $e_1\omega_0$ when $e_1$ is larger than a critical value (e.g., $e_1 \gtrsim 0.7129$ when $R=20$), that is, the QB cannot be further charged in this case. Of course, the critical $e_1$ can be increased by increasing the coupling strength. By comparing the right three panels of Fig. \ref{fig:2}, one can see that for this scenario, $\mathcal{E}_c \equiv 0$ when $e_1=0$, and when $e_1\in(0,1)$, the incoherent and coherent components of the ergotropy $\mathcal{E}$ dominate its dynamics alternately. Similar to that of scenario \uppercase\expandafter{\romannumeral+1}, there is always a moment when $\mathcal{E}$ is zero or infinitesimal. Physically, this phenomenon for both scenarios is due to the fact that the indirect coupling between the charger and the QB is induced by the cavity mode through the exchange of cavity photons, thus there is a competition between the energy flow from the charger to the QB and energy backflow to the charger, both via the mediation of the cavity field. In fact, at the time when $\mathcal{E} \simeq 0$, much energy flows back to the charger and the cavity field. Moreover, as shown in Fig. \ref{fig:2}, there is a series of peaks, and as time goes by, the intensity becomes smaller and smaller. This is understandable as with the evolving of time, more and more energy is dissipated into the cavity field, and in the infinite time limit, the dissipated energy is $c_1\omega_0/2$ for scenario \uppercase\expandafter{\romannumeral+1} and $(1+3e_1)\omega_0/2$ for scenario \uppercase\expandafter{\romannumeral+2}.

\begin{figure}
\centering
\resizebox{0.45 \textwidth}{!}{%
\includegraphics{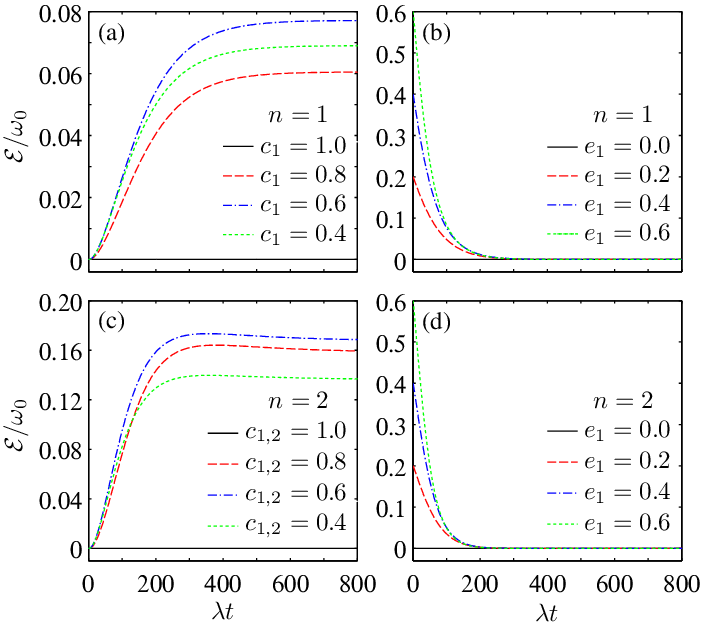}}
\caption{Dynamics of $\mathcal{E}$ of the QB in the bad cavity limit $R=0.1$ for the case of $n$ charging units. The left (right) two panels are plotted with the charger and QB being in the initial states $|\psi\rangle_{c_1\ldots c_n}$ and $|0\rangle$ ($|1\rangle^{\otimes n}$ and $|\varphi\rangle_{e_1}$), respectively.} \label{fig:3}
\end{figure}

In the above we have analyzed charging of a QB in the good cavity limit. Next, we see the case of the bad cavity limit. As shown in Fig. \ref{fig:3}(a), the ergotropy $\mathcal{E}$ for scenario \uppercase\expandafter{\romannumeral+1} increases very slowly with the increase of $\lambda t$, and the asymptotic value is much smaller than $\bar{\mathcal{E}}$ in the good cavity limit. Physically, such an inefficiency has its roots in the weak coupling of the qubits to the reservoir, which induces a very weak indirect interaction between the charger and the QB, thus most of the energy in the charger is leaked into the reservoir. In particular, when $c_1=1$ (i.e., the charger is in its excited state), the QB cannot be charged at all. But, this does not mean that there is no energy being transferred to the QB, as for this case $\rho_{\mathrm{ba}} = \mathrm{diag} \{|\nu_2|^2,1-|\nu_2|^2\}$ (see Appendix \ref{sec:A}), and one can show that $|\nu_2|$ increases from 0 to 0.5 as the time evolves, which yields a nonzero mean energy $E_{\mathrm{ba}}= |\nu_2|^2 \omega_0$ of the QB and a vanishing ergotropy as $\mathcal{E}=\max\{0,2|\nu_2|^2-1\} \equiv 0$. Hence, the energy cannot be extracted by means of any cyclic unitary operation. We have also calculated the incoherent and coherent components of $\mathcal{E}$ and it is found that $\mathcal{E}\equiv \mathcal{E}_c$. For scenario \uppercase\expandafter{\romannumeral+2}, although the initial QB possesses nonzero ergotropy, from Fig. \ref{fig:3}(b) one can see that $\mathcal{E}$ always decays with the evolving time, so the QB cannot be charged in this case.

\begin{figure}
\centering
\resizebox{0.45 \textwidth}{!}{%
\includegraphics{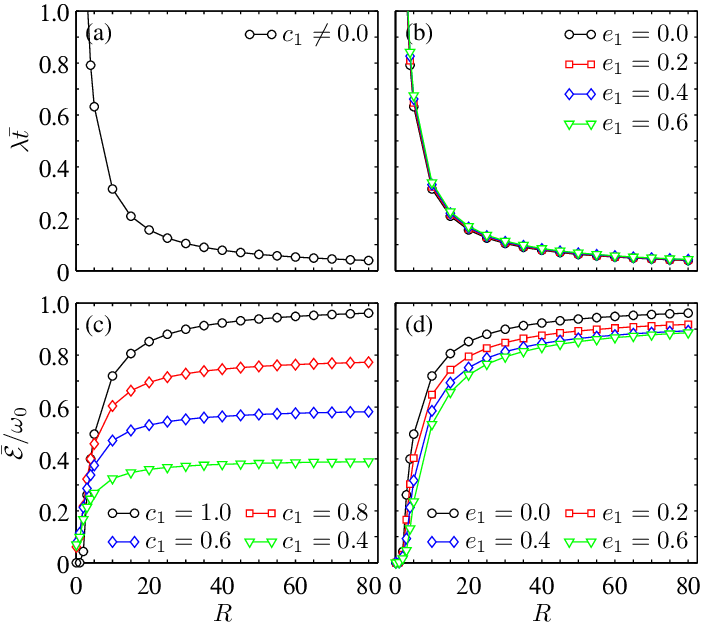}}
\caption{The $R$ dependence of the charging time $\lambda\bar{t}$ and ergotropy $\bar{\mathcal{E}}$ charged by a single charging unit at time $\bar{t}$. The left (right) two panels are plotted with the charger and QB being in the initial states $|\psi\rangle_{c_1}$ and $|0\rangle$ ($|1\rangle$ and $|\varphi\rangle_{e_1}$), respectively. For small $R$, $\lambda\bar{t}$ is very large and the top two panels are cut to better show its behavior in the large $R$ region, and the first five points in (c) and (d) are plotted with $R=0.1$, 1, 2, 3, and 4, respectively.} \label{fig:4}
\end{figure}

In the following, we examine the charging rate of the QB characterized by the charging time $\bar{t}$. As is shown in the top two panels of Fig. \ref{fig:4}, for scenario \uppercase\expandafter{\romannumeral+1}, $\lambda\bar{t}$ is independent of $c_1 \neq 0$ and it decreases monotonically with the increase of $R$. In the bad cavity limit, $\lambda\bar{t}$ will be infinitely large. On the contrary, it is very small in the good cavity limit (e.g., $\lambda\bar{t} \simeq 0.0314$ for $R=100$), that is, a fast charging is achieved. In fact, for scenario \uppercase\expandafter{\romannumeral+1}, both the ergotropy charged on the QB and the charging time can be obtained analytically (see Appendix \ref{sec:A}); specifically, $\bar{t} =2\pi/|\zeta|$ with $\zeta=\lambda(1-4R^2)^{1/2}$. Hence, it is the same for different $c_1\neq 0$. For scenario \uppercase\expandafter{\romannumeral+2}, $\bar{t}$ is weakly dependent on $e_1$, and for a given $e_1 \neq 1$ it also decreases monotonically with the increase of $R$. This shows that the charging rate of a QB could be efficiently enhanced by increasing the coupling strength between the charger-battery system and the reservoir. Moreover, from Figs. \ref{fig:4}(a) and (b) one can note that the charging times for scenarios \uppercase\expandafter{\romannumeral+1} and \uppercase\expandafter{\romannumeral+2} are approximately the same, that is, the residual ergotropy in the QB does not help to improve the charging rate. For scenario \uppercase\expandafter{\romannumeral+2}, however, it is hard to obtain an analytical solution of the ergotropy $\mathcal{E}$. Physically, the insensitivity of the charging time $\bar{t}$ to the initial charger and battery states is due to the fact that the indirect coupling of the charger and QB is induced by the cavity mode through the exchange of cavity photons, and it is the effective coupling strength $\Omega$ rather than the initial system state that dominates the cavity losses (photon escape rate) \cite{common1,common2}, thus it is understandable that $\bar{t}$ is insensitive to the initial system state. The decrease of $\bar{t}$ with the increasing $R$ can also be explained from this perspective as $R=\sqrt{2}\Omega/\lambda$. The ergotropy charged on the QB strongly depends on the initial charger state $|\psi\rangle_{c_1}$, the reason for which is that different charger states have different amounts of internal energy.

In the bottom two panels of Fig. \ref{fig:4}, we show the $R$ dependence of the ergotropy $\bar{\mathcal{E}}$ charged on the QB at $t=\bar{t}$. One can see that for both scenarios \uppercase\expandafter{\romannumeral+1} and \uppercase\expandafter{\romannumeral+2}, $\bar{\mathcal{E}}$ increases monotonically with the increase of $R$. When $R \rightarrow \infty$, $\bar{\mathcal{E}}$ approaches its asymptotic value, which, for scenario \uppercase\expandafter{\romannumeral+1}, equals $c_1 \omega_0$, and for scenario \uppercase\expandafter{\romannumeral+2}, equals $\omega_0$. This indicates that in theory, the energy in the charger is almost fully transferred to extractable work in the QB for such a limiting case. In fact, if the charger is in the excited state and QB is in the ground state, even for the $R\simeq 10$ case which is experimentally accessible \cite{apl}, the QB can be charged up to $\bar{\mathcal{E}}\simeq 0.72 \omega_0$.

\begin{figure}
\centering
\resizebox{0.45 \textwidth}{!}{%
\includegraphics{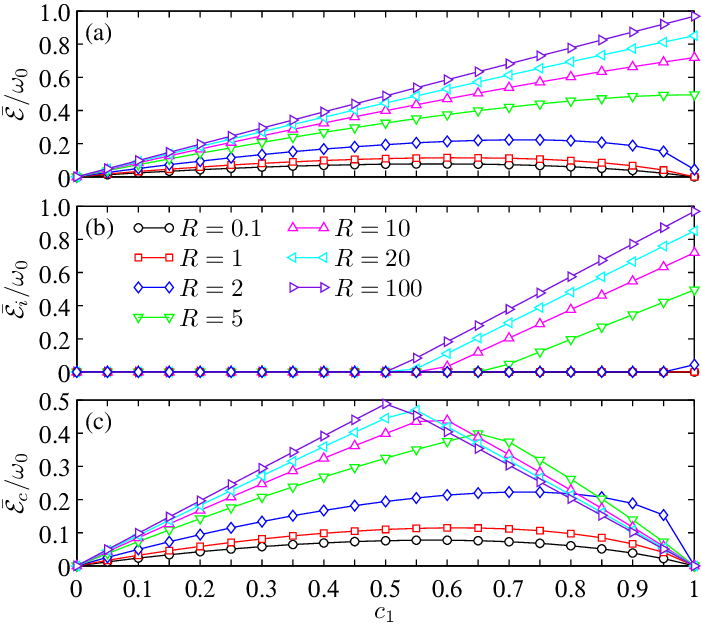}}
\caption{The ergotropy $\bar{\mathcal{E}}$ and its incoherent component $\bar{\mathcal{E}}_i$ and coherent component $\bar{\mathcal{E}}_c$ (in units of $\omega_0$) charged on the QB at time $\bar{t}$ for the case of a single charging unit. The charger and QB are in the initial states $|\psi\rangle_{c_1}$ and $|0\rangle$, respectively. The circles and squares in (b) overlap with each other in the whole parameter region of $c_1$.} \label{fig:5}
\end{figure}

\begin{figure}
\centering
\resizebox{0.45 \textwidth}{!}{%
\includegraphics{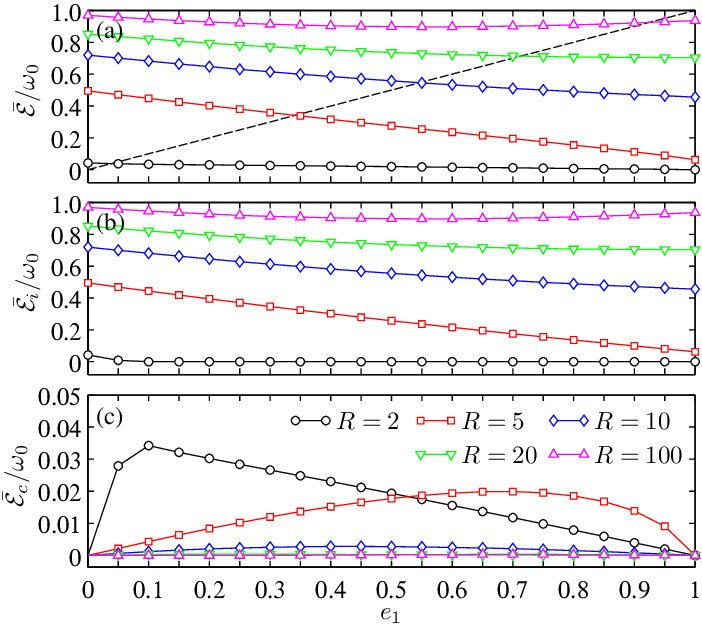}}
\caption{The ergotropy $\bar{\mathcal{E}}$ and its incoherent component $\bar{\mathcal{E}}_i$ and coherent component $\bar{\mathcal{E}}_c$ (in units of $\omega_0$) charged on the QB at time $\bar{t}$ for the case of a single charging unit. The charger and QB are in the initial states $|1\rangle$ and $|\varphi\rangle_{e_1}$, respectively. In the region above the dashed line shown in (a), $\bar{\mathcal{E}}$ exceeds the initial ergotropy $e_1 \omega_0$ in the QB.} \label{fig:6}
\end{figure}

When $R$ is finite, the ergotropy $\bar{\mathcal{E}}$ charged on a QB depends on $c_1$ and $e_1$. To elucidate this in detail, we show in Fig. \ref{fig:5} the $c_1$ dependence of $\bar{\mathcal{E}}$ with the QB being in its ground state initially, and in Fig. \ref{fig:6} the $e_1$ dependence of $\bar{\mathcal{E}}$ with the charger being in the excited state initially. For scenario \uppercase\expandafter{\romannumeral+1}, when $c_1=0$, $\bar{\mathcal{E}}=0$ by definition. When $c_1$ increases from 0 to 1, as depicted in Fig. \ref{fig:5}(a), $\bar{\mathcal{E}}$ first increases to a small peak and then decreases to zero (e.g., $R=0.1$ and $1$) or to a small but finite value (e.g., $R=2$) in the weak coupling regime. In the strong coupling regime, however, $\bar{\mathcal{E}}$ increases monotonically from 0 to its maximum; in particular, for large enough $R$ (e.g., $R=100$), it increases nearly linearly with the increase of $c_1$. Moreover, by comparing the three panels in Fig. \ref{fig:5}, one can also see that the coherent contribution to ergotropy is dominant when $c_1$ is relatively small, while this is not the case for large enough $c_1$, and this phenomenon is consistent with that shown in Fig. \ref{fig:2}. In particular, if the charger is in the excited state (i.e., $c_1=1$) initially, $\bar{\mathcal{E}} \equiv \bar{\mathcal{E}}_i$. This is because for this case, the battery state $\rho_{\mathrm{ba}}(\bar{t})$ at time $\bar{t}$ is diagonal, thereby there is no coherent contribution to the ergotropy \cite{shihl2}.

For scenario \uppercase\expandafter{\romannumeral+2}, the QB is not fully discharged at the initial time, that is, there is residual ergotropy in it. For this case, when $R$ is very small (i.e., the bad cavity limit), the QB cannot be charged, so we plot in Fig. \ref{fig:6} the case of $R\geqslant 2$, from which one can see that even for $R=2$, the charged ergotropy $\bar{\mathcal{E}}$ is still very small, and it exceeds the initial ergotropy $e_1 \omega_0$ of the QB only in a very narrow region of $e_1$ (i.e., $e_1\lesssim 0.0382$). By increasing the coupling strength of the qubits and reservoir, $\bar{\mathcal{E}}$ can be noticeably enhanced. From Fig. \ref{fig:6}(a) one can also see that the region of $e_1$ in which $\bar{\mathcal{E}}$ is larger than the initial ergotropy $e_1\omega_0$ expands with increasing $R$. For example, for $R=100$, the associated region can be obtained approximately as $e_1 \lesssim 0.9239$. As for the ratio of the incoherent and coherent components in $\bar{\mathcal{E}}$, from the bottom two panels of Fig. \ref{fig:6}, one can see that apart from the small $R$ case (e.g., $R=2$), the incoherent contribution is always dominant, and this is different from that for scenario \uppercase\expandafter{\romannumeral+1}. We would also like to mention that $\bar{\mathcal{E}}$ shown in Fig. \ref{fig:6}(a) is not always a monotonic decreasing function of $e_1$ (or equivalently, the initial ergotropy in the QB), e.g., when $R=100$ it takes a minimum at $e_1 \simeq 0.5636$. Nonetheless, $\bar{\mathcal{E}}$ always takes its maximum at $e_1=0$, that is, the QB is in its ground state and empty initially. We have also examined the mixed battery state $\rho_{\mathrm{ba}}=\mathrm{diag}\{e_1, 1-e_1\}$, which is empty for $e_1 \leqslant 0.5$, although its energy is $e_1\omega_0$. It is found that the charging performance is less efficient than that prepared in the ground state. In this sense, the optimal charging of a QB is achieved starting from its ground state, which is in agreement with the findings of Ref. \cite{initial}.

Up to now, we have elucidated how the enlarged $R$ helps to accelerate the charging rate and to pump more ergotropy (i.e., the maximal amount of extractable work) in the QB. Here, it is natural to ask another intriguing question: does the mean energy or ergotropy of the charger determine the ergotropy charged on a QB? As they always coexist for the free Hamiltonian $H_{\mathrm{ch}}$ [see Eq. \eqref{eq3-2}] and the initial state $|\psi\rangle_{c_1}$ [see Eq. \eqref{eq4-1}], we turn to consider a general one-qubit charger state $\rho_{\mathrm{ch}}$ and we first figure out the condition under which there is nonzero energy and zero ergotropy in it. After some algebra, one can obtain that this is achieved for $\rho_{\mathrm{ch}}=\mathrm{diag}\{c_1,1-c_1\}$ with $c_1\leqslant 0.5$. In this case, the mean energy $c_1\omega_0$ in the charger equals that for the initial state $|\psi\rangle_{c_1}$, and in the good cavity limit, most of the energy can be pumped in the QB. However, it is found that for any initial state (including the mixed one) of the QB, there is no ergotropy being charged on it. In this sense, it seems that it is the charger ergotropy instead of its energy that determines the ergotropy charged on a QB.

To gain more insight into the wireless charging process of the QB, we can further consider the dynamics of the mean energy $E_{\mathrm{ch}}$ and ergotropy $\mathcal{E}_{\mathrm{ch}}$ in the charger. Physically, as there is no direct interaction between the charger and the QB, the energy (ergotropy) in the charger will first be pumped in the reservoir and then charged into the QB via the reservoir-mediated indirect interaction. During such a dynamical process, part of the energy (ergotropy) will inevitably be lost. Here, we focus on the case of not very small $R$ and calculate the ergotropy in the charger with the same parameters as in Fig. \ref{fig:2} (for the conciseness of this paper, we do not show them here). For scenario \uppercase\expandafter{\romannumeral+1}, it is found that the ergotropy of the charger reaches its first dynamical minimum at $t=\bar{t}$, and such a minimum decreases with the increase of $R$, for example, for $R=20$ it is of approximately 0. For scenario \uppercase\expandafter{\romannumeral+2}, however, the ergotropy $\mathcal{E}$ in the QB does not always increase synchronously with the decrease of the ergotropy $\mathcal{E}_{\mathrm{ch}}$ in the charger. Specifically, the ergotropy for both the charger and the QB decays to zero after a short period of evolution time and then turns to be increased to the first dynamical maxima. However, the former decays to zero faster than the latter. This shows that there is a delayed effect for the ergotropy pumped in the reservoir to be charged into the QB.

\subsection{The case of two charging units} \label{sec:4b}
In this subsection, we consider the case of two charging units, that is, the total number of qubits $N=3$, where the first two qubits are treated as the charger (i.e., $n=2$) and the third one as a QB. Intuitively, in this case the charger may produce a better result in ergotropy performance than that with a single charging unit as it possesses more energy and ergotropy. But is that really the case? This is our motivation for considering this problem. In particular, the presence of three qubits in the common reservoir makes the charging mechanism complex \cite{workext1,shihl1,shihl2}. In the following, we investigate it in detail for the two scenarios of charging sketched in Fig. \ref{fig:1}.

\begin{figure}
\centering
\resizebox{0.45 \textwidth}{!}{%
\includegraphics{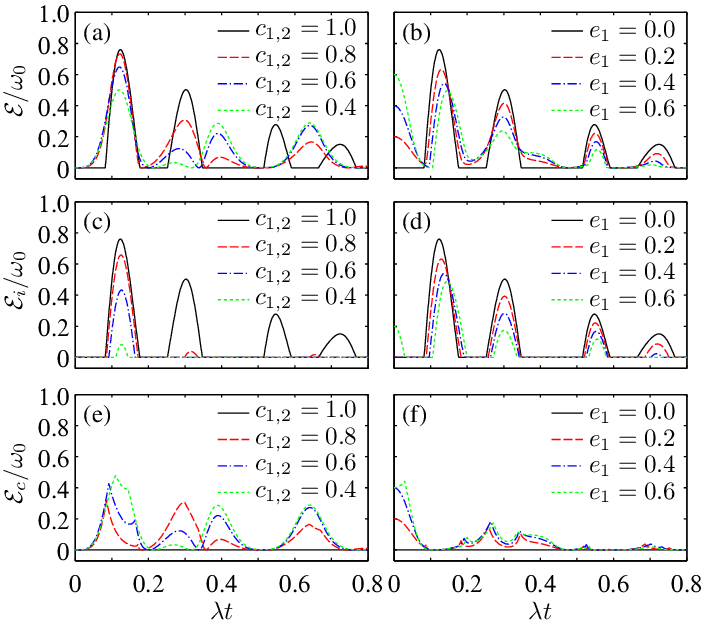}}
\caption{Dynamics of $\mathcal{E}$, $\mathcal{E}_i$, and $\mathcal{E}_c$ of the QB in the good cavity limit $R=20$ for the case of two charging units. The left (right) three panels are plotted with the charger and QB being in the initial states $|\psi\rangle_{c_1 c_2}$ and $|0\rangle$ ($|11\rangle$ and $|\varphi\rangle_{e_1}$), respectively.} \label{fig:7}
\end{figure}

First, we show in Fig. \ref{fig:7} the time evolution of $\mathcal{E}$, $\mathcal{E}_i$, and $\mathcal{E}_c$ of the QB in the good cavity limit $R=20$, where we have chosen $c_1=c_2$ and the values of the system parameters are the same as those in Fig. \ref{fig:2}. From this figure one can note that for both scenarios \uppercase\expandafter{\romannumeral+1} and \uppercase\expandafter{\romannumeral+2}, $\mathcal{E}$ shows a structurally similar behavior to that with a single charging unit. Moreover, although in this case the initial energy and ergotropy in the charger equal twice those for a single charging unit, the dynamical maximum $\bar{\mathcal{E}}$ is slightly decreased. This implies that the amounts of initial energy and ergotropy in the charger are not the only ingredients determining the ergotropy charged on a QB.

The incoherent and coherent components of ergotropy also show similar behaviors to those obtained for the case of a single charging unit shown in Fig. \ref{fig:2}. Specifically, except the case of $c_1=1$ and $e_1=0$ for which $\mathcal{E}_c \equiv 0$, for scenario \uppercase\expandafter{\romannumeral+1} (\uppercase\expandafter{\romannumeral+2}) with $c_1\gtrsim 0.5078$ ($e_1 \neq 0$ and 1), the two components of ergotropy dominate its dynamics alternately, whereas for scenario \uppercase\expandafter{\romannumeral+1} with $c_1 \lesssim 0.5078$, the coherent contribution is always dominant. From Fig. \ref{fig:7} one can also see that both $\mathcal{E}$ and $\mathcal{E}_i$ reach their first dynamical maxima at the same interaction time $\bar{t}$, but this is  not the case for $\mathcal{E}_c$ in general.

In the bad cavity limit $R=0.1$, as exemplified in the bottom two panels of Fig. \ref{fig:3}, the ergotropy $\mathcal{E}$ also shows qualitatively the same behavior as that for the case of a single charging unit. But for scenario \uppercase\expandafter{\romannumeral+1}, apart from the special case of $c_{1,2}=1$ for which $\mathcal{E}$ always remains zero, the dynamical maximum of $\mathcal{E}$ in this case is noticeably enhanced, and this phenomenon is in sharp contrast to that in the good cavity limit.

\begin{figure}
\centering
\resizebox{0.45 \textwidth}{!}{%
\includegraphics{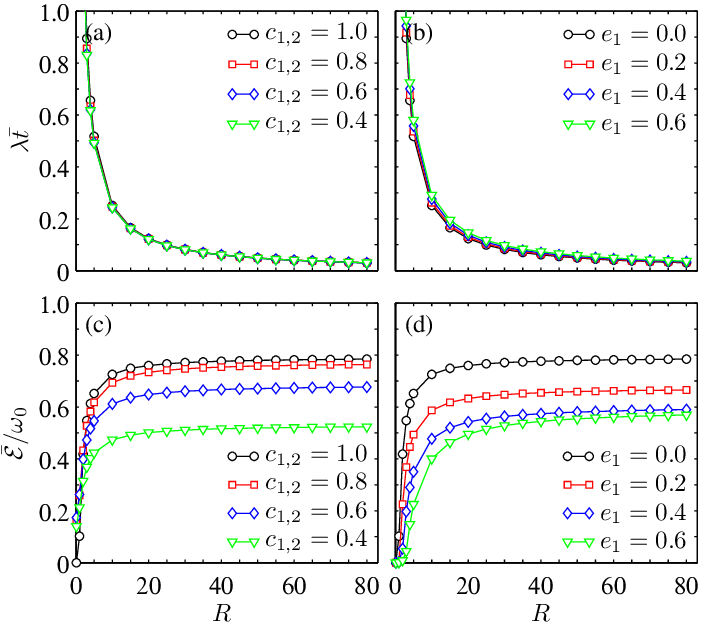}}
\caption{The $R$ dependence of the charging time $\lambda\bar{t}$ and ergotropy $\bar{\mathcal{E}}$ charged by two charging units at time $\bar{t}$ . The left (right) two panels are plotted with the charger and QB being in the initial states $|\psi\rangle_{c_1 c_2}$ and $|0\rangle$ ($|11\rangle$ and $|\varphi\rangle_{e_1}$), respectively. For small $R$, $\lambda\bar{t}$ is very large and the top two panels are cut to better see its behavior in the large $R$ region, and the first five points in (c) and (d) are plotted with $R=0.1$, 1, 2, 3, and 4, respectively.} \label{fig:8}
\end{figure}

In Fig. \ref{fig:8}, we show the $R$ dependence of the charging time $\bar{t}$ and the associated $\bar{\mathcal{E}}$ with different system parameters. First, one can see that its behavior is very similar to that for a single charging unit (cf. Fig. \ref{fig:4}). Specifically, it is also insensitive to the variation of $c_{1,2}$ (scenario \uppercase\expandafter{\romannumeral+1}) and $e_1$ (scenario \uppercase\expandafter{\romannumeral+2}), especially in the strong coupling regime. This is also due to the fact that the indirect coupling of the charger and QB is induced by the cavity mode through the exchange of cavity photons, thus the decay rate of the pseudomode is determined by the coupling strength $\Omega$ (note that $R=\sqrt{2}\Omega/\lambda$) \cite{common1,common2}. By comparing Figs. \ref{fig:4} and \ref{fig:8}, one can also find that $\bar{t}$ is slightly shortened in this case. Moreover, while $\bar{\mathcal{E}}$ also approaches an asymptotic value when $R\rightarrow \infty$, in contrast to the case of a single charging unit, such an asymptotic value is smaller than the initial energy $2c_1\omega_0$ (equals the initial ergotropy) in the two charging units. In particular, for scenario \uppercase\expandafter{\romannumeral+2} with $e_1 \gtrsim 0.5911$ [see, e.g., the triangles in Fig. \ref{fig:8}(d)], $\bar{\mathcal{E}}$ even cannot exceed its initial value $e_1\omega_0$, that is, the QB in this case cannot be charged anymore by the two charging units. This indicates that in the good cavity limit, the two charging units cannot outperform the single charging one if there is a considerable amount of residual ergotropy in the QB initially. Of course, the case may be different for the finite $R$, and we will discuss this issue in detail after introducing the three charging units.

\begin{figure}
\centering
\resizebox{0.45 \textwidth}{!}{%
\includegraphics{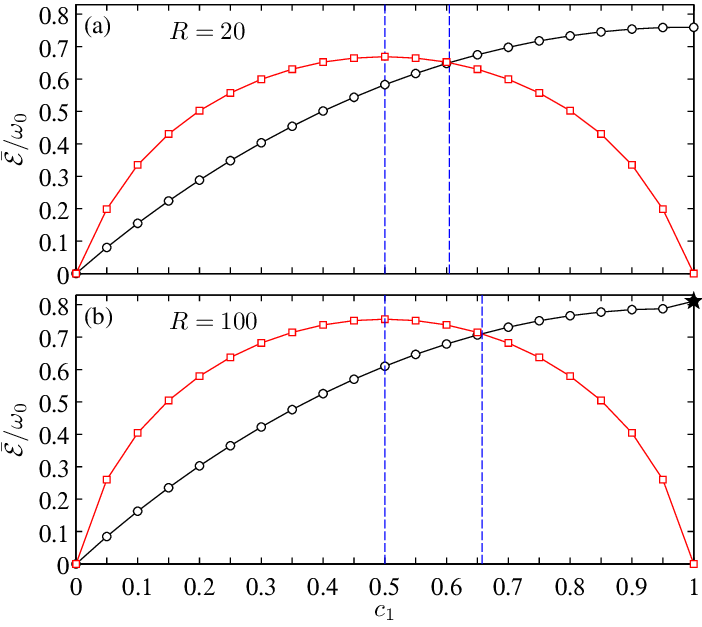}}
\caption{Comparison of $\bar{\mathcal{E}}$ charged on a QB in the ground state initially, where the charger is in the initial state $|\psi\rangle_{c_1 c_1}$ (circles) or $|\Psi^{+}\rangle$ (squares). The left dashed lines correspond to $c_1=0.5$, after which the charger energy $E_{\mathrm{ch}}(|\Psi^{+}\rangle) < E_{\mathrm{ch}}(|\psi\rangle_{c_1 c_1})$, while the right dashed lines at $c_1 \simeq 0.6045$ (a) and $0.6575$ (b), respectively, show the critical points at which $\bar{\mathcal{E}}$ for $|\Psi^{+}\rangle$ equals that for $|\psi\rangle_{c_1 c_1}$. The star in (b) denotes the case for which $\bar{\mathcal{E}}$ corresponds to the second instead of the first dynamical maximum of $\mathcal{E}$.} \label{fig:9}
\end{figure}

As we studied in the above only the initial product charger states, it is natural to ask whether the charging performance of a QB can be improved by using the initially correlated charger state. To answer this question, we consider the following Bell-like charger states:
\begin{equation}\label{eq4-2}
\begin{aligned}
 & |\Psi^{\pm}\rangle = \sqrt{c_1}|10\rangle \pm \sqrt{1-c_1}|01\rangle, \\
 & |\Phi^{\pm}\rangle = \sqrt{c_1}|11\rangle \pm \sqrt{1-c_1}|00\rangle,
\end{aligned}
\end{equation}
which have the same entanglement. In Fig. \ref{fig:9}, we give a plot of $\bar{\mathcal{E}}$ charged on a QB in its ground state by using the charger in the initial states $|\psi\rangle_{c_1 c_1}$ and $|\Psi^{+}\rangle$, respectively. The initial charger energy is $E_{\mathrm{ch}}(|\psi\rangle_{c_1 c_1})= 2 c_1\omega_0$ and $E_{\mathrm{ch}}(|\Psi^{+}\rangle)=\omega_0$ ($\forall c_1$), respectively, and so is the initial charger ergotropy. As a result, $E_{\mathrm{ch}}(|\Psi^{+}\rangle) < E_{\mathrm{ch}}(|\psi\rangle_{c_1 c_1})$ when $c_1 > 0.5$. As is shown in Fig. \ref{fig:9}, the charged ergotropy with the initial charger state $|\Psi^{+}\rangle$ decreases with the increasing $|c_1-0.5|$, which is proportional to the initial entanglement in it. Moreover, the curves of ergotropy $\bar{\mathcal{E}}$ for these two initial charger states cross each other at a critical point $c_{1,r}$. In the region between 0 and $c_{1,r}$, $\bar{\mathcal{E}}$ for the initial charger state $|\Psi^{+}\rangle$ is larger than that for the initial charger state $|\psi\rangle_{c_1 c_1}$; in particular, it is worth emphasizing that when $c_1 \in (0.5,c_{1,r})$, the mean energy in $|\Psi^{+}\rangle$ is less than that in $|\psi\rangle_{c_1 c_1}$. This indicates that, under certain situations, the correlated charger states are more efficient than that of the uncorrelated ones. But, this is not a universal conclusion. For example, $\bar{\mathcal{E}} \equiv 0$ for the initial charger state $|\Psi^{-}\rangle$. This is understandable as it is robust against the decoherence effect; in particular, it reduces to the decoherence-free state when $c_1=0.5$ \cite{common1,common2}, for which there is no energy being transferred to the QB. For the initial charger states $|\Phi^\pm\rangle$, a further calculation reveals that they are also less efficient than $|\psi\rangle_{c_1 c_1}$, although they have the same entanglement with $|\Psi^+\rangle$ and there is a considerable amount of energy ($E_{\mathrm{ba}}\simeq 0.8799 c_1 \omega_0$) being transferred to the QB. Thus, the advantage of the entangled states depends on the structure of the entanglement, the spectral properties of the reservoir, and the coupling regime. Moreover, as highlighted by the star in Fig. \ref{fig:9}(b), the maximum ergotropy $\bar{\mathcal{E}}$ charged on the QB may correspond to the second instead of the first dynamical maximum of $\mathcal{E}$.

We have also examined time evolution of the ergotropy $\mathcal{E}_{\mathrm{ch}}$ in the two charging units, and for the conciseness of this paper, we also do not display them here. Note that $\mathcal{E}_{\mathrm{ch}}$ represents the ergotropy of the charger with respect to the total state $\rho_{\mathrm{ch}}$ of the two charging units, which is larger than the sum of the ergotropies with respect to the reduced states of the two charger qubits \cite{EPL}. Similar to that for a single charging unit, the interaction time at which $\mathcal{E}_{\mathrm{ch}}$ reaches its first dynamical minimum is also not synchronous with the charging time $\bar{t}$ at which the ergotropy charged on the QB reaches its dynamical maximum. Specifically, for scenario \uppercase\expandafter{\romannumeral+1} with not very small $R$, $\mathcal{E}_{\mathrm{ch}}$ reaches its first dynamical minimum after $t=\bar{t}$ if $c_{1,2}$ is very large (e.g., $c_{1,2}\gtrsim 0.9756$ when $R=20$), while for the other $c_{1,2}$, it reaches its first dynamical minimum before $t=\bar{t}$. For scenario \uppercase\expandafter{\romannumeral+2} with not very small $R$, however, $\mathcal{E}_{\mathrm{ch}}$ reaches its first dynamical minimum after $t=\bar{t}$ for a very small $e_1$ (e.g., $e_1 \lesssim 0.0476$ when $R=20$), and otherwise, it reaches its first dynamical minimum before $t=\bar{t}$. This indicates that the energy pumped in the reservoir previously can still be transferred to the QB in the time interval during which there is no additional energy to be injected into the reservoir.

\subsection{The case of three charging units}\label{sec:4c}
When three qubits in the reservoir are treated as the charger ($n=3$) and the fourth one as a QB, the behaviors of $\mathcal{E}$, $\mathcal{E}_i$, and $\mathcal{E}_c$ are also structurally similar to those showed in Figs. \ref{fig:2} and \ref{fig:7}, so we do not display them here. The charging time $ \bar{t}$ and the ergotropy $\bar{\mathcal{E}}$ charged on the QB at $t=\bar{t}$ also show qualitatively a similar $R$ dependence to those showed in Figs. \ref{fig:4} and \ref{fig:8}, so we also do not display them here. We only point out here that compared to those achieved with the two charging units, $\lambda \bar{t}$ for both scenarios \uppercase\expandafter{\romannumeral+1} and \uppercase\expandafter{\romannumeral+2} can be further slightly shortened, while $\bar{\mathcal{E}}$ can also be enhanced to some extent.

\begin{figure}
\centering
\resizebox{0.45 \textwidth}{!}{%
\includegraphics{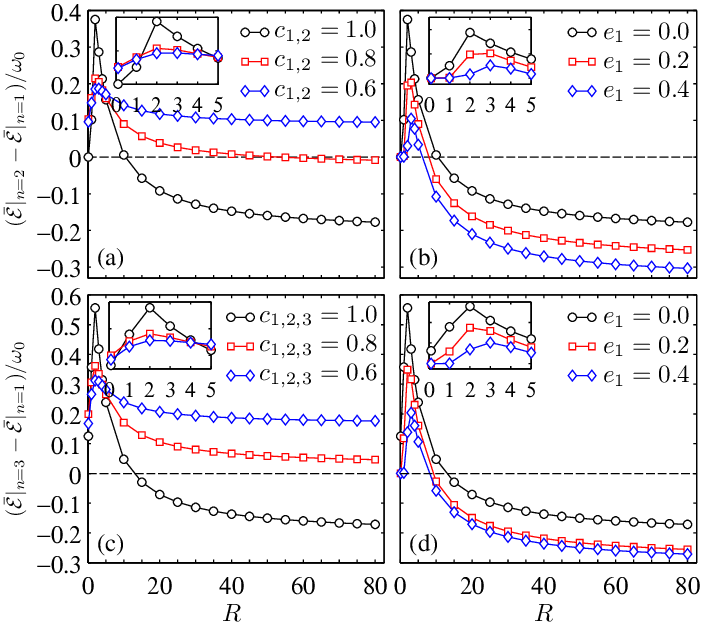}}
\caption{Comparison of $\bar{\mathcal{E}}$ charged on the QB with different number $n$ of charging units. The left (right) two panels are plotted with the chargers and QB being in the initial states $|\psi\rangle_{c_1\ldots c_n}$ and $|0\rangle$ ($|1\rangle^{\otimes n}$ and $|\varphi\rangle_{e_1}$), respectively. The first five points in each panel are plotted with $R=0.1$, 1, 2, 3, and 4, respectively.} \label{fig:10}
\end{figure}

Before ending this section, we provide some further comparison about the charging of a QB by using different numbers of charging units. First, we present in Fig. \ref{fig:10} a comparison of the ergotropy, where we have denoted by $\bar{\mathcal{E}}_{n=1}$ the ergotropy charged on a QB with $n=1$, and likewise for $\bar{\mathcal{E}}_{n=2}$ and $\bar{\mathcal{E}}_{n=3}$. For scenario \uppercase\expandafter{\romannumeral+1}, from Fig. \ref{fig:10}(a) one can see that if $c_{1,2}=1$, $\bar{\mathcal{E}}_{n=2} > \bar{\mathcal{E}}_{n=1}$ when $R \lesssim 10.34$, and this region expands with a decrease in $c_{1,2}$. This, together with Fig. \ref{fig:10}(c), shows that if the charger energy is not maximal initially, one can enhance the ergotropy of the QB by using multiple charging units. As for scenario \uppercase\expandafter{\romannumeral+2}, from Fig. \ref{fig:10}(b) one can observe that the region in which $\bar{\mathcal{E}}_{n=2} > \bar{\mathcal{E}}_{n=1}$ shrinks with the increase of $e_1$. Although Fig. \ref{fig:10}(d) shows that the ergotropy can be slightly enhanced for $n=3$, the enhancement is limited. So in contrast to scenario \uppercase\expandafter{\romannumeral+1}, the performance of a QB can be enhanced by using multiple charging units only in the not very strong coupling regime. We have also compared the ergotropy charged on the initially mixed battery state $\rho_{\mathrm{ba}}=\mathrm{diag}\{e_1, 1-e_1\}$ with different numbers of charging units, and a qualitatively similar phenomenon to that showed in Fig. \ref{fig:10} is observed.

\begin{figure}
\centering
\resizebox{0.45 \textwidth}{!}{%
\includegraphics{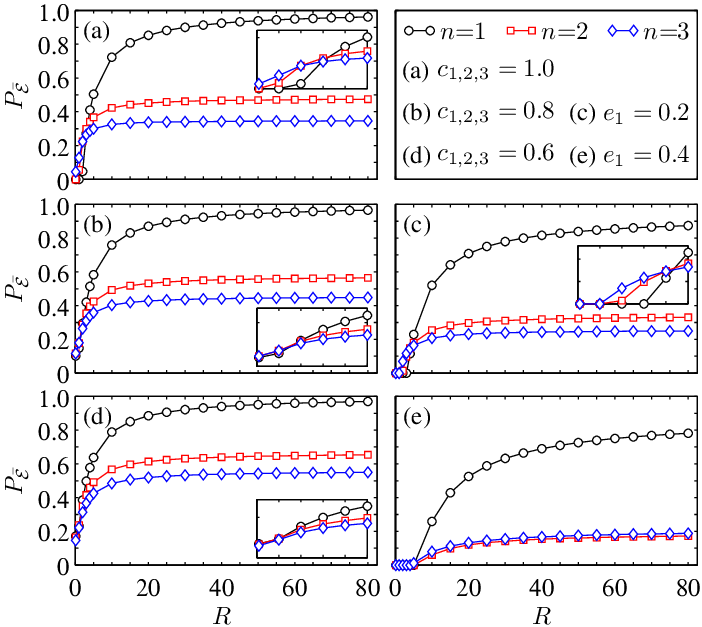}}
\caption{Comparison of the charging efficiency $P_{\bar{\mathcal{E}}}$ for different number $n$ of charging units. The left (right) panels are plotted with the charger and QB being in the initial states $|\psi\rangle_{c_1\ldots c_n}$ and $|0\rangle$ ($|1\rangle^{\otimes n}$ and $|\varphi\rangle_{e_1}$), respectively. The parameters are given in the top right corner, and the insets show $P_{\bar{\mathcal{E}}}$ for $R=0.1$, $1$, $2$, $3$, $4$, and $5$, respectively.} \label{fig:11}
\end{figure}

One might also be concerned with the charging efficiency, that is, the proportion of output energy of the charger that can be converted into extractable work in the QB. To answer this question, we define
\begin{equation}\label{eq4-3}
 P_{\bar{\mathcal{E}}}= \frac{\Delta\bar{\mathcal{E}}}{\Delta E_{\mathrm{ch}}}
                      = \frac{\max\{0,\bar{\mathcal{E}}(\bar{t})-\bar{\mathcal{E}}(0)\}}
                        {\tr\{[\rho_{\mathrm{ch}}(0)-\rho_{\mathrm{ch}}(\bar{t})]H_{\mathrm{ch}}\}},
\end{equation}
where $\Delta\bar{\mathcal{E}}$ is the net ergotropy charged on the QB, $\Delta E_{\mathrm{ch}}$ is the energy output from the charger, while $\rho_{\mathrm{ch}}(\bar{t})$ denotes the state of the charger at time $\bar{t}$, and likewise for $\rho_{\mathrm{ch}}(0)$. Based on this definition we performed numerical calculations and the result is shown in Figs. \ref{fig:11}. Clearly, $P_{\bar{\mathcal{E}}}$ can be noticeably enhanced by increasing the coupling strength. In particular, for scenario \uppercase\expandafter{\romannumeral+1} with a single charging unit, if the QB is in its ground state initially (cf. the circles in the left column of Fig. \ref{fig:11}), $P_{\bar{\mathcal{E}}}$ approaches 1 for large enough $R$. Moreover, for the case of not very weak coupling, $P_{\bar{\mathcal{E}}}$ decreases when $n$ increases from 1 to 3, whereas this may not always be the case in the weak coupling regime. For scenario \uppercase\expandafter{\romannumeral+2}, as depicted in Figs. \ref{fig:11}(c) and (e), if a single charging unit is considered, $P_{\bar{\mathcal{E}}}$ also approaches 1 when $R$ approaches infinity, but in this case its increase with the increase of $R$ is slower than that for scenario \uppercase\expandafter{\romannumeral+1}. When the two and three charging units are considered, although $P_{\bar{\mathcal{E}}}$ also increases with the increase of $R$, the asymptotic value at $R \rightarrow \infty$ is obviously less than 1. This implies that a considerable amount of energy output from the charger is leaked into the reservoir in this situation.

Apart from $P_{\bar{\mathcal{E}}}$ in Eq. \eqref{eq4-3}, we can also consider other definitions of the charging efficiency, e.g.,
\begin{equation}\label{eq4-4}
 \mathcal{P}_{\bar{\mathcal{E}}}= \frac{\Delta\bar{\mathcal{E}}}{\Delta E_{\mathrm{ba}}}
                                = \frac{\max\{0,\bar{\mathcal{E}}(\bar{t})-\bar{\mathcal{E}}(0)\}}
                                  {\tr\{[\rho_{\mathrm{ba}}(\bar{t})-\rho_{\mathrm{ba}}(0)]H_{\mathrm{ch}}\}},
\end{equation}
which is the proportion of input energy that can be extracted via the optimal cyclic unitary transformation. As its behavior is qualitatively similar to that in Fig. \ref{fig:11} and no new physics can be obtained, we do not plot them here.

\section{Summary and discussion} \label{sec:5}
In summary, we have studied the wireless charging of a QB via $n$ ($n\geqslant 1$) charging units, where both the charger and the QB interact with a structured bosonic reservoir (which may be implemented, e.g., by an electromagnetic field inside a lossy cavity) and there is no direct coupling between them. Inspired by the fact that in a realistic situation, the charger energy may be not maximal and one rarely waits until a battery runs out before charging, we considered two scenarios of wireless charging, that is, scenario \uppercase\expandafter{\romannumeral+1} in which the charger is in a superposition of the excited and ground states and the QB is empty (i.e., fully discharged) initially, and scenario \uppercase\expandafter{\romannumeral+2} in which the charger is in its excited state and the QB has residual ergotropy (i.e., partially discharged) initially. For both scenarios, we considered the ergotropy, which quantifies the maximum amount of work a QB could supply during unitary cycles, as a figure of merit for comparing the charging performance. Our results showed that the charging time $\bar{t}$, defined as the interaction time at which the QB reaches its dynamical maximum, decreases with the increase of the coupling strength. The charging time $\bar{t}$ also slightly shortens with an increase in the number of the charging units. The ergotropy $\bar{\mathcal{E}}$ charged on the QB at $t=\bar{t}$, on the other hand, increases with the increasing coupling strength. Specifically, if a single charging unit is used, then in the strong coupling regime, the ergotropy $\bar{\mathcal{E}}$ for scenario \uppercase\expandafter{\romannumeral+1} increases with the increasing amount of the initial energy in the charger, while that for scenario \uppercase\expandafter{\romannumeral+2} is not very sensitive to the residual ergotropy in the QB. In particular, the QB is almost fully charged when the coupling strength is strong enough. In the weak and moderate coupling regimes, however, the case will be different. The incoherent and coherent contributions to ergotropy are also different in different coupling regimes. Moreover, compared to the case of a single charging unit, the ergotropy $\bar{\mathcal{E}}$ charged by two and three charging units is decreased in the strong coupling regime, whereas in the weak and moderate coupling regimes, it can be enhanced to some extent; in particular, such an enhancement is pronounced for scenario \uppercase\expandafter{\romannumeral+1} in which the charger energy is not maximal, and this implies that in this situation the multiple charging units could compensate to some extent the lack of energy in a single charging unit. Finally, we have also considered the charging efficiency which quantifies the proportion of output energy that can be converted into extractable work in the QB. It was found that it also increases with the increase of the coupling strength. If a single charging unit is considered, this efficiency approaches 1 when $R\rightarrow\infty$, while for that of the two and three charging units, it is significantly decreased.

While we considered in this paper the charging of a single-cell QB via $n$ charging units, one may also be concerned with the general case in which an $m$-cell QB is charged via $n$ charging units. Although it is hard to obtain a general result for this case due to the limited computing resource, we performed numerical calculations with $n+m\leqslant 4$ and the results showed that for $m=2$ and 3, the dynamical behavior of $\mathcal{E}$ is structurally similar to that for a single-cell QB, and the maximum ergotropy $\bar{\mathcal{E}}$ charged on a QB can also be noticeably enhanced by increasing the coupling strength, but now it cannot be fully charged, even if $R$ approaches infinity. Moreover, similar to that shown in Fig. \ref{fig:9}(b), if the charger and QB are in the initial states $|11\rangle$ and $|00\rangle$ (i.e., $n=2$ and $m=2$), respectively, the maximum ergotropy $\bar{\mathcal{E}}$ also corresponds to the second instead of the first dynamical maximum of $\mathcal{E}$ when $R\gtrsim 34$. It is also interesting that the ergotropy $\bar{\mathcal{E}}$ charged on a two-cell QB is much larger than that charged on a single-cell QB when using the same two charging units.

\begin{figure}
\centering
\resizebox{0.45 \textwidth}{!}{%
\includegraphics{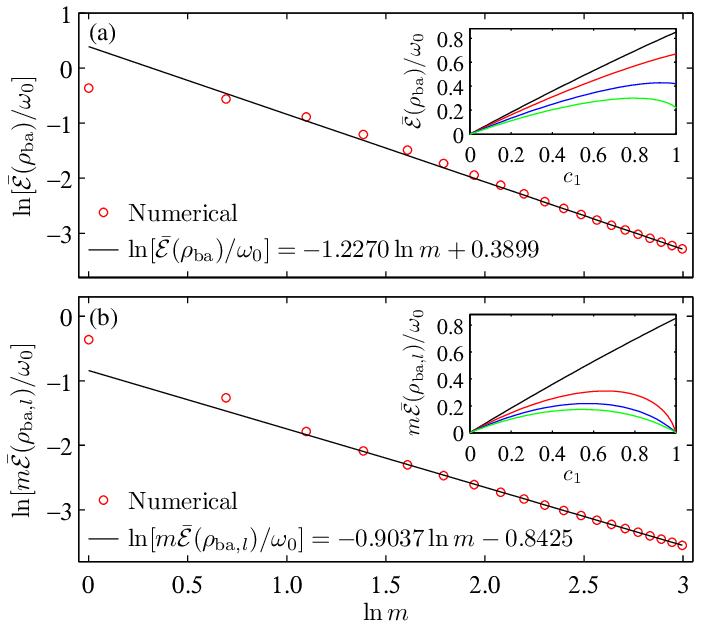}}
\caption{Scaling behaviors of $\bar{\mathcal{E}}(\rho_\mathrm{ba})$ and $m\bar{\mathcal{E}}(\rho_{\mathrm{ba},l})$ in terms of the number $m$ of the battery cell, where $R=20$, $c_1=0.8$, and the charger and QB are in the initial states $|\psi\rangle_{c_1}$ and $|0\rangle^{\otimes m}$, respectively. The four lines from top to bottom in each inset are plotted with $m=1$, 2, 3, and 4, respectively.} \label{fig:12}
\end{figure}

For the special case of the initial state $|\psi\rangle_{c_1} \otimes |0\rangle^{\otimes m}$, that is, the single charger unit is in the state $|\psi\rangle_{c_1}$ of Eq. \eqref{eq4-1} and the $m$-cell QB is in the state $|0\rangle^{\otimes m}$ initially, one can obtain analytically the ergotropy $\mathcal{E}(\rho_\mathrm{ba})$ charged on the $m$-cell QB and the ergotropy $\mathcal{E}(\rho_{\mathrm{ba},l})$ charged on each battery cell (see Appendix \ref{sec:A}). Based on these analytical results, one can show that in the strong coupling regime, for any fixed $m$, the charging time $\bar{t}$ is the same for different $c_1 \neq 0$, and it scales as $\ln (\lambda \bar{t}) \sim -0.5\ln (m+1) +\ln (\sqrt{2}\pi/R)$. Moreover, as is shown in Fig. \ref{fig:12}(a), except for the cases of $m=1$ and 2, the maximal ergotropy $\bar{\mathcal{E}}(\rho_\mathrm{ba})$ charged on the $m$-cell QB is not a monotonic function of $c_1$. As for $m \bar{\mathcal{E}}(\rho_{\mathrm{ba},l})$, from Fig. \ref{fig:12}(b) one can note that it is also not a monotonic function of $c_1$ when $m \geqslant 2$. In addition, one always has $\bar{\mathcal{E}}(\rho_\mathrm{ba}) > m \bar{\mathcal{E}}(\rho_{\mathrm{ba},l})$ when $m \geqslant 2$, and such a collective effect is rooted in the fact that the product of $m$ passive states may not always be passive \cite{entangleu}. Finally, as can be seen from Fig. \ref{fig:12}, for any $c_1 \neq 0$ ($c_1 \neq 0$ and 1), $\bar{\mathcal{E}}(\rho_\mathrm{ba})$ [$m\bar{\mathcal{E}}(\rho_{\mathrm{ba},l})$] decreases with increasing $m$, and in the large $m$ region, $\ln \bar{\mathcal{E}}(\rho_\mathrm{ba})$ and $\ln [m \bar{\mathcal{E}}(\rho_{\mathrm{ba},l})]$ are found to be scaled with different exponents. As a result, if the number of battery cells is very large, the QB cannot be charged by a single charging unit. In the weak coupling regime, an analysis similar to that for $R=20$ shows that the ergotropy also shows a similar scaling behavior, so we do not explicitly present the plots here for the purpose of brevity.

The above results suggest that, in general, the charging process of a QB is extremely complex. It depends on the initial states of the charger and the QB, as well as the coupling manner and coupling strength between them. To find more intrinsic relations underlying these elements and to identify essential ingredients boosting the charging performance of a QB is an intriguing direction. This will not only help us to have a better understanding of the charging mechanism, but can also help us to develop the high-efficiency QB.

Moreover, it is natural for future work to consider the case where the charger and the QB are coupled to a common reservoir with different strengths, just as the two-qubit case studied in Ref. \cite{wireless1}, or further consider the case where there is frequency detuning between the qubits and the cavity field, for which the off-resonant effect on generation of two-qubit entanglement has been studied \cite{common1,common2}. In addition, one can also generalize the results in this work to the structured reservoir, which acts as a mediator between the charger and QB, with other kinds of spectra (e.g., the sub-Ohmic, Ohmic, and super-Ohmic spectral densities \cite{Ohmic}). Of course, in all these cases we first need to solve the dynamical equation of the system for the general instead of a special multiqubit initial state, which may be a very intricate and challenging task. Beyond that, it would also be worthwhile to further consider the finite temperature reservoir and explore how it affects the wireless charging of a QB. Intuitively, the increasing temperature may degrade the charging performance as the thermal fluctuations will induce more energy dissipation. But a general study on the details of the finite temperature effects is still needed. Finally, while we take a multimode cavity field as the mediator for energy exchange, another direction for future study may be the Tavis-Cummings model, where the atoms play the role of both charger and battery, and a single-mode cavity field plays the mediated role in the charging process. Similar studies on the Tavis-Cummings QB can be found in Refs. \cite{qcqb1,shihl3,TCqb1,noiseqb8}, where the cavity field in various initial states (Fock state, coherent state, squeezed vacuum state, etc.) serves as the charger and $N$ noninteracting two-level atoms serve as the QB.

\section*{ACKNOWLEDGMENTS}
This work was supported by the National Natural Science Foundation of China (Grants No. 12275212, No. T2121001, No. 92265207, and No. 92365301), Shaanxi Fundamental Science Research Project for Mathematics and Physics (Grant No. 22JSY008), the Youth Innovation Team of Shaanxi Universities (Grant No. 24JP177), and Technology Innovation Guidance Special Fund of Shaanxi Province (Grant No. 2024QY-SZX-17).

\section*{DATA AVAILABILITY}
The data that support the findings of this article are openly available \cite{data}.

\begin{appendix}
\section{Solution of the model for a special initial state} \label{sec:A}
\setcounter{equation}{0}
\renewcommand{\theequation}{A\arabic{equation}}

When the charger is in the initial state $|\psi\rangle_{c_1}$ of Eq. \eqref{eq4-1} and the $m$-cell QB is in the initial state $|0\rangle^{\otimes m}$, as $|0\rangle^{\otimes (m+1)}$ does not decay in time, the state of the total system (the charger-battery system plus the reservoir) at time $t$ can be written as
\begin{equation}\label{eqa-1}
\begin{aligned}
 |\Psi(t)\rangle= \, & \sqrt{c_1} \Bigg[\nu_1(t) |1\rangle_S |\bar{0}\rangle_R
                        + \nu_2(t) \sum_{j=2}^{m+1} |j\rangle_S |\bar{0}\rangle_R \\
                     &  + \sum_k \nu_k(t)|0\rangle_S |\bar{1}_k\rangle_R \Bigg]
                        + \sqrt{1-c_1}|0\rangle_S|\bar{0}\rangle_R,
\end{aligned}
\end{equation}
where $|0\rangle_S=|0\rangle^{\otimes (m+1)}$ denotes the ground state of the charger-battery system, $|\bar{0}\rangle_R$ is the vacuum state of the reservoir, $|j\rangle_S$ is the state of the charger-battery system with only qubit $j$ being in the excited state, and $|\bar{1}_k\rangle_R$ is the state of the reservoir with only one excitation in mode $k$. Using the same method as in Ref. \cite{common1}, we obtain the coefficients $\nu_1(t)$ and $\nu_2(t)$ as
\begin{equation}\label{eqa-2}
 \nu_1(t)=\frac{p(t)+m}{m+1},~
 \nu_2(t)=\frac{p(t)-1}{m+1},
\end{equation}
where the time-dependent parameter $p(t)$ is given by
\begin{equation}\label{eqa-3}
 p(t)=e^{-\frac{1}{2}\lambda t}\left(\cosh\frac{\zeta t}{2}+\frac{\lambda}{\zeta} \sinh\frac{\zeta t}{2}\right),
\end{equation}
with $\zeta= \lambda\sqrt{1-2(m+1)R^2}$.

From Eq. \eqref{eqa-1}, one can obtain the state of the $m$-cell QB as
\begin{equation}\label{eqa-4}
 \rho_\mathrm{ba}= |\xi\rangle \langle\xi| + c_1\big(1-m|\nu_2|^2\big) |0\rangle_\mathrm{ba}\langle 0|,
\end{equation}
where
\begin{equation}\label{eqa-5}
  |\xi\rangle= \sqrt{c_1}\nu_2 \sum_{l=1}^m |l\rangle_\mathrm{ba} + \sqrt{1-c_1}|0\rangle_\mathrm{ba},
\end{equation}
with $|0\rangle_\mathrm{ba}$ and $|l\rangle_\mathrm{ba}$ being analogous to $|0\rangle_S$ and $|j\rangle_S$, respectively, and the only difference is that the former one is for the charger-battery system and the latter one is for the QB. Then, the ergotropy in the $m$-cell QB can be obtained as
\begin{equation}\label{eqa-6}
\begin{aligned}
 \mathcal{E}(\rho_\mathrm{ba}) = \, & \omega_0 \Bigg [m c_1|\nu_2|^2 \\
                                    & +\frac{1}{2} \sqrt{1+4m c_1^2|\nu_2|^2(m|\nu_2|^2-1)}- \frac{1}{2}\Bigg].
\end{aligned}
\end{equation}

Similarly, the state $\rho_{\mathrm{ba},l}$ for the battery cell $l$ ($l=1,\dots,m$) can be obtained as
\begin{equation}\label{eqa-7}
 \rho_{\mathrm{ba},l}=
 \begin{pmatrix}
   c_1|\nu_2|^2            & \sqrt{c_1(1-c_1)}\nu_2 \vspace{0.5em} \\
   \sqrt{c_1(1-c_1)}\nu^*_2  & 1-c_1|\nu_2|^2
 \end{pmatrix},
\end{equation}
which is the same for different $l$. Hence, the ergotropy in each battery cell can be obtained as
\begin{equation}\label{eqa-8}
 \mathcal{E}(\rho_{\mathrm{ba},l}) = \omega_0 \Bigg[c_1|\nu_2|^2
                                     +\frac{1}{2}\sqrt{1+4c_1^2|\nu_2|^2(|\nu_2|^2-1)} - \frac{1}{2} \Bigg].
\end{equation}

From Eqs. \eqref{eqa-6} and \eqref{eqa-8}, one can show that if $2(m+1)R^2>1$, then for any $c_1 \neq 0$, $\partial \mathcal{E}(\rho)/\partial |\nu_2| \geqslant 0$ ($\rho\in\{\rho_\mathrm{ba},\rho_{\mathrm{ba},l}\}$), that is, $\mathcal{E}(\rho)$ is always a nondecreasing function of $|\nu_2|$. Thus, it takes its maximum when $|\nu_2|$ takes its maximum. Based on such an observation, one can obtain the charging time $\bar{t}$ as
\begin{equation}\label{eqa-9}
  \bar{t}= 2\pi/|\zeta|,
\end{equation}
which is the same for both $\mathcal{E}(\rho_\mathrm{ba})$ and $\mathcal{E}(\rho_{\mathrm{ba},l})$. In the strong coupling region, $\bar{t}$ can be approximated as $\bar{t}\simeq \pi/[(m+1)^{1/2}\Omega]$. By substituting $\bar{t}$ into Eqs. \eqref{eqa-6} and \eqref{eqa-8}, we obtain the maximal ergotropies $\bar{\mathcal{E}}(\rho_\mathrm{ba})$ charged on the $m$-cell QB and $\bar{\mathcal{E}}(\rho_{\mathrm{ba},l})$ charged on each battery cell, respectively.

If $2(m+1)R^2<1$, both $\mathcal{E}(\rho_\mathrm{ba})$ and $\mathcal{E}(\rho_{\mathrm{ba},l})$ take their maxima at $t\rightarrow \infty$, hence the charging time is $\bar{t} \rightarrow \infty$. This corresponds to $p(\bar{t})=0$ and $\nu_2(\bar{t})=-1/(m+1)$, and substituting these into Eqs. \eqref{eqa-6} and \eqref{eqa-8}, we obtain the corresponding $\bar{\mathcal{E}}(\rho_\mathrm{ba})$ and $\bar{\mathcal{E}}(\rho_{\mathrm{ba},l})$, respectively.

\end{appendix}

\newcommand{\PRL}{Phys. Rev. Lett. }
\newcommand{\RMP}{Rev. Mod. Phys. }
\newcommand{\PRA}{Phys. Rev. A }
\newcommand{\PRB}{Phys. Rev. B }
\newcommand{\PRD}{Phys. Rev. D }
\newcommand{\PRE}{Phys. Rev. E }
\newcommand{\PRX}{Phys. Rev. X }
\newcommand{\APL}{Appl. Phys. Lett. }
\newcommand{\NJP}{New J. Phys. }
\newcommand{\JPA}{J. Phys. A }
\newcommand{\JPB}{J. Phys. B }
\newcommand{\PLA}{Phys. Lett. A }
\newcommand{\NP}{Nat. Phys. }
\newcommand{\NC}{Nat. Commun. }
\newcommand{\EPL}{Europhys. Lett. }
\newcommand{\AdP}{Ann. Phys. (Berlin) }
\newcommand{\AoP}{Ann. Phys. (N.Y.) }
\newcommand{\QIP}{Quantum Inf. Process. }
\newcommand{\PR}{Phys. Rep. }
\newcommand{\SR}{Sci. Rep. }
\newcommand{\JMP}{J. Math. Phys. }
\newcommand{\RPP}{Rep. Prog. Phys. }
\newcommand{\PA}{Physica A }
\newcommand{\CMP}{Commun. Math. Phys. }
\newcommand{\SCPMA}{Sci. China-Phys. Mech. Astron. }


\begin{thebibliography}{50}
\bibitem{EPL} A. E. Allahverdyan, R. Balian, and Th. M. Nieuwenhuizen, Maximal work extraction from finite quantum systems, \EPL \textbf{67}, 565 (2004).
\bibitem{passive1} W. Pusz and S. L. Woronowicz, Passive states and KMS states for general quantum systems, Commun. Math. Phys. \textbf{58}, 273 (1978).
\bibitem{passive2} A. Lenard, Thermodynamical proof of the Gibbs formula for elementary quantum systems, J. Stat. Phys. \textbf{19}, 575 (1978).
\bibitem{passive3} M. Alimuddin, T. Guha, and P. Parashar, Structure of passive states and its implication in charging quantum batteries, \PRE \textbf{102}, 022106 (2020).
\bibitem{entangleu} R. Alicki and M. Fannes, Entanglement boost for extractable work from ensembles of quantum batteries, \PRE \textbf{87}, 042123 (2013).
\bibitem{workext1} K. V. Hovhannisyan, M. Perarnau-Llobet, M. Huber, and A. Ac\'{\i}n, Entanglement generation is not necessary for optimal work extraction, \PRL \textbf{111}, 240401 (2013).
\bibitem{workext2} M. Perarnau-Llobet, K. V. Hovhannisyan, M. Huber, P. Skrzypczyk, N. Brunner, and A. Ac\'{\i}n, Extractable work from correlations, \PRX \textbf{5}, 041011 (2015).
\bibitem{qcqb0} A. Mukherjee, A. Roy, S. S. Bhattacharya, and M. Banik, Presence of quantum correlations results in a nonvanishing ergotropic gap, \PRE \textbf{93}, 052140 (2016).
\bibitem{qcqb1} G. M. Andolina, M. Keck, A. Mari, M. Campisi, V. Giovannetti, and M. Polini, Extratable work, the role of correlations, and asymoptotic freedom in quantum batteries, \PRL \textbf{122}, 047702 (2019).
\bibitem{qcqb2} L. P. Garc\'{\i}a-Pintos, A. Hamma, and A. del Campo, Fluctuations in extractable work bound the charging power of quantum batteries, \PRL \textbf{125}, 040601 (2020).

\bibitem{qcqb3} G. Francica, F. C. Binder, G. Guarnieri, M. T. Mitchison, J. Goold, and F. Plastina, Quantum coherence and ergotropy, \PRL \textbf{125}, 180603 (2020).
\bibitem{qcqb4} B. \c{C}akmak, Ergotropy from coherences in an open quantum system, \PRE \textbf{102}, 042111 (2020).
\bibitem{qcqb5} F. H. Kamin, F. T. Tabesh, and S. Salimi, Entanglement, coherence, and charging process of quantum batteries, \PRE \textbf{102}, 052109 (2020).
\bibitem{shihl1} J. X. Liu, H. L. Shi, Y. H. Shi, X. H. Wang, and W. L. Yang, Entanglement and work extraction in the central-spin quantum battery, \PRB \textbf{104}, 245418 (2021).
\bibitem{qcqb6} G. Francica, Quantum correlations and ergotropy, \PRE \textbf{105}, L052101 (2022).
\bibitem{correch} M. B. Arjmandi, A. Shokri, E. Faizi, and H. Mohammadi, Performance of quantum batteries with correlated and uncorrelated chargers, \PRA \textbf{106}, 062609 (2022).
\bibitem{shihl2} H. L. Shi, S. Ding, Q. K. Wan, X. H. Wang, and W. L. Yang, Entanglement, coherence, and extractable work in quantum batteries, \PRL \textbf{129}, 130602 (2022).
\bibitem{luomx} X. Yang, Y. H. Yang, M. Alimuddin, R. Salvia, S. M. Fei, L. M. Zhao, S. Nimmrichter, and M. X. Luo, Battery capacity of energy-storing quantum systems, \PRL \textbf{131}, 030402 (2023).
\bibitem{shihl3} H. Y. Yang, H. L. Shi, Q. K. Wan, K. Zhang, X. H. Wang, and W. L. Yang, Optimal energy storage in the Tavis-Cummings quantum battery, \PRA \textbf{109}, 012204 (2024).
\bibitem{AVS} J. Y. Gyhm and U. R. Fischer, Beneficial and detrimental entanglement for quantum battery charging, AVS Quantum Sci. \textbf{6}, 012001 (2024).

\bibitem{spin_qb1} T. P. Le, J. Levinsen, K. Modi, M. M. Parish, and F. A. Pollock, Spin-chain model of a many-body quantum battery, \PRA \textbf{97}, 022106 (2018).
\bibitem{spin_qb2} D. Rossini, G. M. Andolina, and M. Polini, Many-body localized quantum batteries, \PRB \textbf{100}, 115142 (2019).
\bibitem{spin_qb3} S. Ghosh, T. Chanda, and A. Sen(De), Enhancement in the performance of a quantum battery by ordered and disordered interactions, \PRA \textbf{101}, 032115 (2020).
\bibitem{spin_qb4} F. Q. Dou, H. Zhou, and J. A. Sun, Cavity Heisenberg-spin-chain quantum battery, \PRA \textbf{106}, 032212 (2022).
\bibitem{spin_qb5} R. Grazi, D. Sacco Shaikh, M. Sassetti, N. Traverso Ziani, and D. Ferraro, Controlling energy storage crossing quantum phase transitions in an integrable spin quantum battery, \PRL \textbf{133}, 197001 (2024).
\bibitem{TCqb1} W. Lu, J. Chen, L. M. Kuang, and X. Wang, Optimal state for a Tavis-Cummings quantum battery via the Bethe ansatz method, \PRA \textbf{104}, 043706 (2021).
\bibitem{noiseqb0} D. Farina, G. M. Andolina, A. Mari, M. Polini, and V. Giovannetti, Charger-mediated energy transfer for quantum batteries: an open-system approach, \PRB \textbf{99}, 035421 (2019).
\bibitem{oscillator_qb1} G. M. Andolina, D. Farina, A. Mari, V. Pellegrini, V. Giovannetti, and M. Polini, Charger-mediated energy transfer in exactly solvable models for quantum batteries, \PRB \textbf{98}, 205423 (2018).
\bibitem{oscillator_qb2} Y. V. de Almeida, T. F. F. Santos, and M. F. Santos, Cooperative isentropic charging of hybrid quantum batteries, \PRA \textbf{108}, 052218 (2023).
\bibitem{Dickeqb1} L. Fusco, M. Paternostro, and G. De Chiara, Work extraction and energy storage in the Dicke model, \PRE \textbf{94}, 052122 (2016).

\bibitem{Dickeqb2} D. Ferraro, M. Campisi, G. M. Andolina, V. Pellegrini, and M. Polini, High-power collective charging of a solid-state quantum battery, \PRL \textbf{120}, 117702 (2018).
\bibitem{Dickeqb3} Y. Y. Zhang, T. R. Yang, L. Fu, and X. Wang, Powerful harmonic charging in a quantum battery, \PRE \textbf{99}, 052106 (2019).
\bibitem{Dickeqb4} F. Q. Dou, Y. Q. Lu, Y. J. Wang, and J. A. Sun, Extended Dicke quantum battery with interatomic interactions and driving field, \PRB \textbf{105}, 115405 (2022).
\bibitem{threelevelqb} A. C. Santos, B. \c{C}akmak, S. Campbell, and N. T. Zinner, Stable adiabatic quantum batteries, \PRE \textbf{100}, 032107 (2019).
\bibitem{Dou1} D. L. Yang, F. M. Yang, and F. Q. Dou, Three-level Dicke quantum battery, \PRB \textbf{109}, 235432 (2024).
\bibitem{Dou2} F. M. Yang and F. Q. Dou, Resonator-qutrit quantum battery, \PRA \textbf{109}, 062432 (2024).
\bibitem{chpf0} F. Campaioli, F. A. Pollock, F. C. Binder, L. C\'{e}leri, J. Goold, S. Vinjanampathy, and K. Modi, Enhancing the charging power of quantum batteries, \PRL \textbf{118}, 150601 (2017).
\bibitem{chpf1} F. Pirmoradian and K. M{\o}lmer, Aging of a quantum battery, \PRA \textbf{100}, 043833 (2019).
\bibitem{thermalization} K. V. Hovhannisyan, F. Barra, and A. Imparato, Charging assisted by thermalization, Phys. Rev. Research \textbf{2}, 033413 (2020).
\bibitem{chpf2}A. C. Santos, A. Saguia, and M. S. Sarandy, Stable and charge-switchable quantum batteries, \PRE \textbf{101}, 062114 (2020).

\bibitem{chpf3} L. Peng, W. B. He, S. Chesi, H. Q. Lin, and X. W. Guan, Lower and upper bounds of quantum battery power in multiple central spin systems, \PRA \textbf{103}, 052220 (2021).
\bibitem{chpf4} S. Ghosh and A. Sen(De), Dimensional enhancements in a quantum battery with imperfections, \PRA \textbf{105}, 022628 (2022).
\bibitem{chpf5} A. Crescente, M. Carrega, M. Sassetti, and D. Ferraro, Ultrafast charging in a two-photon Dicke quantum battery, \PRB \textbf{102}, 245407 (2020).
\bibitem{chpf6} M. B. Arjmandi, H. Mohammadi, A. Saguia, M. S. Sarandy, and A. C. Santos, Localization effects in disordered quantum batteries, \PRE \textbf{108}, 064106 (2023).
\bibitem{chpf7} P. Chen, T. S. Yin, Z. Q. Jiang, and G. R. Jin, Quantum enhancement of a single quantum battery by repeated interactions with large spins, \PRE \textbf{106}, 054119 (2022).
\bibitem{Shangc} Z. G. Lu, G. Tian, X. Y. L\"{u}, and C. Shang, Topological quantum batteries, arXiv:2405.03675.
\bibitem{advan1} F. C. Binder, S. Vinjanampathy, K. Modi, and J. Goold, Quantacell: powerful charging of quantum batteries, \NJP \textbf{17}, 075015 (2015).
\bibitem{advan2} G. M. Andolina, M. Keck, A. Mari, V. Giovannetti, and M. Polini, Quantum versus classical many-body batteries, \PRB \textbf{99}, 205437 (2019).
\bibitem{SYKbattery} D. Rossini, G. M. Andolina, D. Rosa, M. Carrega, and M. Polini, Quantum advantage in the charging process of Sachdev-Ye-Kitaev batteries, \PRL \textbf{125}, 236402 (2020).
\bibitem{advan3} A. C. Santos, Quantum advantage of two-level batteries in the self-discharging process, \PRE \textbf{103}, 042118 (2021).


\bibitem{advan4} S. Seah, M. Perarnau-Llobet, G. Haack, N. Brunner, and S. Nimmrichter, Quantum speed-up in collisional battery charging, \PRL \textbf{127}, 100601 (2021).
\bibitem{advan5} J. Y. Gyhm, D. \v{S}afr\'{a}nek, and D. Rosa, Quantum charging advantage cannot be extensive without global operations, \PRL \textbf{128}, 140501 (2022).
\bibitem{advan6} T. F. F. Santos, Y. V. de Almeida, and M. F. Santos, Vacuum-enhanced charging of a quantum battery, \PRA \textbf{107}, 032203 (2023).
\bibitem{advan7} R. Salvia, M. Perarnau-Llobet, G. Haack, N. Brunner, and S. Nimmrichter, Quantum advantage in charging cavity and spin batteries by repeated interactions, Phys. Rev. Research \textbf{5}, 013155 (2023).
\bibitem{bound1} S. Juli\`{a}-Farr\'{e}, T. Salamon, A. Riera, M. N. Bera, and M. Lewenstein, Bounds on the capacity and power of quantum batteries, Phys. Rev. Research \textbf{2}, 023113 (2020).
\bibitem{bound2} S. Zakavati, F. T. Tabesh, and S. Salimi, Bounds on charging power of open quantum batteries, \PRE \textbf{104}, 054117 (2021).
\bibitem{noiseqb1} F. Barra, Dissipative charging of a quantum battery, \PRL \textbf{122}, 210601 (2019).
\bibitem{wireless1} F. T. Tabesh, F. H. Kamin, and S. Salimi, Environment-mediated charging process of quantum batteries, \PRA \textbf{102}, 052223 (2020).
\bibitem{wireless2} F. H. Kamin, F. T. Tabesh, S. Salimi, F. Kheirandish, and A. C. Santos, Non-Markovian effects on charging and self-discharging process of quantum batteries, \NJP \textbf{22}, 083007 (2020).
\bibitem{noiseqb2} M. Carrega, A. Crescente, D. Ferraro, and M. Sassetti, Dissipative dynamics of an open quantum battery, \NJP \textbf{22}, 083085 (2020).


\bibitem{PRAppl} J. Q. Quach and W. J. Munro, Using dark states to charge and stabilize open quantum batteries, Phys. Rev. Appl. \textbf{14}, 024092 (2020).
\bibitem{noiseqb3} W. Chang, T. R. Yang, H. Dong, L. Fu, X. Wang, and Y. Y. Zhang, Optimal building block of multipartite quantum battery in the driven-dissipative charging, \NJP \textbf{23}, 103026 (2021).
\bibitem{noiseqb4} S. Ghosh, T. Chanda, S. Mal, and A. Sen(De),  Fast charging of a quantum battery assisted by noise, \PRA \textbf{104}, 032207 (2021).
\bibitem{noiseqb5} M. L. Song, L. J. Li, X. K. Song, L. Ye, and D. Wang, Environment-mediated entropic uncertainty in charging quantum batteries, \PRE \textbf{106}, 054107 (2022).
\bibitem{noiseqb6} M. B. Arjmandi, H. Mohammadi, and A. C. Santos, Enhancing self-discharging process with disordered quantum batteries, \PRE \textbf{105}, 054115 (2022).
\bibitem{noiseqb7} K. Xu, H. G. Li, Z. G. Li, H. J. Zhu, G. F. Zhang, and W. M. Liu, Charging performance of quantum batteries in a double-layer environment, \PRA \textbf{106}, 012425 (2022).
\bibitem{collective} F. Mayo and A. J. Roncaglia, Collective effects and quantum coherence in dissipative charging of quantum batteries, \PRA \textbf{105}, 062203 (2022).
\bibitem{noiseqb8} J. Carrasco, J. R. Maze, C. Hermann-Avigliano, and F. Barra, Collective enhancement in dissipative quantum batteries, \PRE \textbf{105}, 064119 (2022).
\bibitem{squeezqb} F. Centrone, L. Mancino, and M. Paternostro, Charging batteries with quantum squeezing, \PRA \textbf{108}, 052213 (2023).
\bibitem{twomode} L. Wang, S. Q. Liu, F. L. Wu, H. Fan, and S. Y. Liu, Two-mode Raman quantum battery dependent on coupling strength, \PRA \textbf{108}, 062402 (2023).

\bibitem{anjh} W. L. Song, H. B. Liu, B. Zhou, W. L. Yang, and J. H. An, Remote charging and degradation suppression for the quantum battery, \PRL \textbf{132}, 090401 (2024).
\bibitem{weak} A. H. A. Malavazi, R. Sagar, B. Ahmadi, and P. R. Dieguez, Weak measurement-based protocol for ergotropy protection in open quantum batteries, arXiv:2411.16633.
\bibitem{initial} A. Crescente, M. Carrega, M. Sassetti, and D. Ferraro, Charging and energy fluctuations of a driven quantum battery, \NJP \textbf{22}, 063057 (2020).
\bibitem{common1} S. Maniscalco, F. Francica, R. L. Zaffino, N. Lo Gullo, and F. Plastina, Protecting entanglement via the quantum Zeno effect, \PRL \textbf{100}, 090503 (2008).
\bibitem{common2} F. Francica, S. Maniscalco, J. Piilo, F. Plastina, and K.-A. Suominen, Off-resonant entanglement generation in a lossy cavity, \PRA \textbf{79}, 032310 (2009).
\bibitem{pmode0} L. Mazzola, S. Maniscalco, J. Piilo, and K.-A. Suominen, Interplay between entanglement and entropy in two-qubit systems, \JPB \textbf{43}, 085505 (2010).
\bibitem{pmode1} B. M. Garraway, Nonperturbative decay of an atomic system in a cavity, \PRA \textbf{55}, 2290 (1997).
\bibitem{pmode2} B. J. Dalton, S. M. Barnett, and B. M. Garraway, Theory of pseudomodes in quantum optical processes, \PRA \textbf{64}, 053813 (2001).
\bibitem{pmode3} B. J. Dalton and B. M. Garraway, Non-Markovian decay of a three-level cascade atom in a structured reservoir, \PRA \textbf{68}, 033809 (2003).
\bibitem{pmode4} L. Mazzola, S. Maniscalco, J. Piilo, K.-A. Suominen, and B. M. Garraway, Sudden death and sudden birth of entanglement in common structured reservoirs, \PRA \textbf{79}, 042302 (2009).

\bibitem{pmode5} D. Z. Rossato, T. Werlang, L. K. Castelano, C. J. Villas-Boas, and F. F. Fanchini, Purity as a witness for initial system-environment correlations in open-system dynamics, \PRA \textbf{84}, 042113 (2011).
\bibitem{apl} S. Kuhr, S. Gleyzes, C. Guerlin, J. Bernu, U. B. Hoff, S. Del\'{e}glise, S. Osnaghi, M. Brune, J.-M. Raimond, S. Haroche, E. Jacques, P. Bosland, and B. Visentin, Ultrahigh finesse Fabry-P\'{e}rot superconducting resonator, \APL \textbf{90}, 164101 (2007).
\bibitem{Ohmic} A. J. Leggett, S. Chakravarty, A. T. Dorsey, M. P. A. Fisher, A. Garg, and W. Zwerger, Dynamics of the dissipative two-state system, \RMP {\bf 59}, 1 (1987).
\bibitem{data} M. L. Hu, T. Gao, and H. Fan, Data for "Efficient wireless charging of a quantum battery" [Data set], Zenodo (2025), doi: \href{https://doi.org/10.5281/zenodo.15162165} {10.5281/zenodo.15162165}.

\end{thebibliography}

\end{document}